\documentclass{aa}
\usepackage{txfonts}
\usepackage{natbib}
\usepackage{amssymb}
\bibpunct{(}{)}{;}{a}{}{,}
\usepackage{epsf}
\usepackage{graphicx}
\usepackage{epsfig}
\usepackage{graphics}
\usepackage{subfigure}
\usepackage[english]{babel}
\usepackage{rotating}
\usepackage{color}

\newcommand{\teff}{$T_{\rm{eff}}$}
\newcommand{\logg}{$\log g$}
\newcommand{\lL}{\ifmmode \log \frac{L}{L_{\sun}} \else $\log \frac{L}{L_{\sun}}$\fi}

\newcommand{\vsini}{$V$~sin$i$}

\newcommand{\vmac}{$v_{\rm mac}$}

\newcommand{\kms}{km~s$^{-1}$}
\newcommand{\msun}{M$_{\sun}$}

\begin{document}

\title{Surface abundances of ON stars\thanks{Based on observations obtained 1) at the Anglo-Australian Telescope; 2) at the Canada-France-Hawaii Telescope (CFHT) which is operated by the National Research Council (NRC) of Canada, the Institut National des Science de l'Univers of the Centre National de la Recherche Scientifique (CNRS) of France, and the University of Hawaii; 3) at the ESO/La Silla Observatory under programs 081.D-2008, 083.D-0589, 086.D-0997; 4) the Nordic Optical Telescope, operated  on  the  island  of  La  Palma  jointly  by  Denmark,  Finland, Iceland, Norway, and Sweden, in the Spanish Observatorio del Roque de  los  Muchachos  of  the  Instituto  de  Astrofísica  de  Canarias; 5) the Mercator Telescope, operated on the island of La Palma by the Flemish Community, at the Spanish Observatorio del Roque de los Muchachos of the Instituto de Astrofísica de Canarias.}}
\author{F. Martins\inst{1}
\and S. Sim\'on-D\'iaz\inst{2}\inst{3}
\and A. Palacios\inst{1}
\and I. Howarth\inst{4}
\and C. Georgy\inst{5}
\and N.R. Walborn\inst{6}
\and J.-C. Bouret\inst{7}
\and R. Barb\'a\inst{8}
}
\institute{LUPM, Universit\'e Montpellier 2, CNRS, Place Eug\`ene Bataillon, F-34095 Montpellier, France  \\
\and
Instituto de Astrofísica de Canarias, 38200, La Laguna, Tenerife, Spain \\
\and
Departamento de Astrof\'isica, Universidad de La Laguna, E-38205 La Laguna, Tenerife, Spain \\
\and
Department of Physics \& Astronomy, University College London, Gower St, London WC1E 6BT, UK \\
\and
Astrophysics group, EPSAM, Keele University, Lennard-Jones Labs, Keele, ST5 5BG, UK \\
\and
Space Telescope Science Institute, 3700 San Martin Drive, Baltimore, MD, 21218, USA \\
\and
Aix Marseille Universit\'e, CNRS, LAM (Laboratoire d'Astrophysique de Marseille) UMR 7326, 13388, Marseille, France \\
\and
Departamento de F\'{\i}sica y Astronom\'{\i}a, Universidad de La Serena, Cisternas 1200 N, La Serena, Chile
}

\offprints{Fabrice Martins\\ \email{fabrice.martins@umontpellier.fr}}

\date{Received / Accepted }

\abstract
{Massive stars burn hydrogen through the CNO cycle during most of their evolution. When mixing is efficient, or when mass transfer in binary systems happens, chemically processed material is observed at the surface of O and B stars. }
{ON stars show stronger lines of nitrogen than morphologically normal
  counterparts. Whether this corresponds to the presence of material processed through the CNO cycle or not is not known. Our goal is to answer this question.}
{We perform a spectroscopic analysis of a sample of ON stars with atmosphere models. We determine the fundamental parameters as well as the He, C, N, and O surface abundances. We also measure the projected rotational velocities. We compare the properties of the ON stars to those of normal O stars.}
{We show that ON stars are usually helium-rich. Their CNO surface abundances are fully consistent with predictions of nucleosynthesis. ON stars are more chemically evolved and rotate - on average - faster than normal O stars. Evolutionary models including rotation cannot account for the extreme enrichment observed among ON main sequence stars. 
 Some ON stars are members of binary systems, but others are single stars as indicated by stable radial velocities. Hence, mass transfer is not a simple explanation for the observed chemical properties.}
{We conclude that ON stars show extreme chemical enrichment at their surface, consistent with nucleosynthesis through the CNO cycle. Its origin is not clear at present.}


\authorrunning{Martins et al.}
\titlerunning{Abundances of ON stars}

\maketitle

\section{Introduction}
\label{s_intro}

Massive stars are born as O and B dwarfs on the main sequence. They subsequently evolve into supergiants of various type (blue, yellow, red) as their effective temperature decreases and their radius increases. Some stars may evolve back and forth between these types of supergiants due to as-yet poorly-known physical mechanisms \citep[e.g.][]{ge14}. Above about 25 \msun\ (at solar metallicity) these stars develop strong winds after the supergiant phase, becoming nitrogen- or carbon-rich Wolf-Rayet stars \citep[WN, WC; e.g.,][]{paul07}. The different types of Wolf-Rayet star reflect the different compositions of their surface material: WN stars show nitrogen enrichments and carbon depletions, while WC stars are hydrogen-free and have a high fraction of carbon in their atmospheres.

These chemical properties are direct consequences of a different evolutionary states. Massive stars burn hydrogen to helium through the partial or complete CNO cycling (depending on the temperature). In equilibrium, most of the nuclei in the CNO cycle are in the form of nitrogen; consequently, nucleo{\-}synthesis in massive stars produces an excess of nitrogen and a depletion of carbon and oxygen. If mixing processes are efficient, part of this processed material can be brought to the surface, and thus be detected spectroscopically. WN stars show the typical chemical patterns of CNO burning; in subsequent evolutionary phases, helium burns to carbon, accounting for the chemical appearance of WC stars. 

The CNO cycle proceeds in the cores of massive stars as long as hydrogen is available (i.e. during the main sequence). Hence we expect massive stars to become increasingly nitrogen rich, and carbon/oxygen poor, as they evolve off the zero-age main sequence.

Rotation is a powerful way of triggering mixing mechanisms in massive stars \citep{mm00,langer12}, and evolutionary calculations of rotating stars show that the surface composition of OB stars can be modified by CNO-processed material, even during early evolutionary phases \citep{brott11a,ek12,cl13}. The degree of enrichment depends on several parameters: rotation speed, metallicity, initial mass, magnetic field. 

Observationally, the predictions of stellar-evolution models incorporating rotation have been partly confirmed. \citet{hunter08,hunter09} showed that the majority of the B~stars they studied in the Galaxy, LMC and SMC exhibit nitrogen enrichments as predicted by rotating models, although a non-negligible fraction (20 to 40\%) were found to be more enriched than expected for their rotation speed. \citet{przy10} and \citet{maeder14} demonstrated that B~stars showed surface CNO patterns consistent with the expectations of nucleo{\-}synthesis; \citet{mimesO} reached similar conclusions for a large sample of Galactic single O stars \citep[see also][]{jc12,jc13}. 

Surface abundances can also be modified by mass transfer in binary systems. If the binary separation is small, the more massive component fills its Roche lobe first, because of its faster evolution, and may dump processed material onto the surface of the secondary \citep{langer08}. The primary may subsequently explode as supernova, which will either disrupt the system, leaving the chemically-contaminated secondary as a single star, or produce a binary system with a high-mass star and a compact object. During the mass-transfer process, the primary may also transfer angular momentum to the secondary, which is thereby spun up \citep{wellstein01,petrovic05}. The faster rotation can trigger additional mixing of CNO material produced in the secondary's core, contributing further to the modification of surface abundances.  Even in the absence of mass transfer, the rotation of binary components may be affected by tidal interactions, with consequences for mixing. 

Surface abundances are therefore a key to understanding the evolution of single and binary massive stars. \citet{walborn70,walborn71,walborn76} reported the existence of O and B stars with peculiar CNO spectra: the OBN and OBC stars \citep[see also][]{walborn04}. In the former, lines of nitrogen (especially \ion{N}{III}~$\lambda\lambda$4630--4640) are much stronger than in normal OB stars, while in the latter, they are weaker. At the same time, \ion{C}{iii}~$\lambda$4650 is weak in OBN stars. Most ON stars have a spectral type between O8.5 and O9.7 -- where \ion{C}{iii} and \ion{N}{iii} lines are easily observed -- but some are also found at spectral type O2, based on the morphology of \ion{N}{iv} and \ion{O}{iv} lines \citep{walborn04}. \citet{lester73} studied the ON star HD~201345 and concluded that its spectroscopic appearance was due to the presence of CNO processed material on its surface. \citet{schon88} studied three ON stars together with two normal O stars, and concluded that the ON stars were helium-rich and showed clear signs of CNO processing at their surfaces. Similar conclusions were reached by \citet{vil02} for the ON star HD~191423. \citet{sh94} determined the helium abundance of one OC, one normal O and one ON star. They found an increasing ratio He/H along the OC/O/ON sequence, confirming the suggestion of \citet{walborn76} that the OBN and OBC stars represent different degrees of chemical evolution of OB stars. \citet{hs01} studied the distribution of rotational velocities of ON stars and concluded that, on average, they rotate faster than normal stars, supporting the hypothesis that rotational mixing could account for their anomalous surface abundances \citep[see also][]{schon88}. \citet{br78} investigated the binary frequency of OBN/OBC stars \citep[see also][]{boy05}. They found that most OBN stars show radial-velocity variations (with some binary systems clearly identified), while OBC stars seem to be constant. 

While these earlier investigations have provided evidence of CNO processing, to date quantitative spectroscopic analyses have focussed on helium \citep{schon88,sh94,vil02}, employing relatively simple model atmospheres; determinations of CNO abundances have relied on equivalent-width measurements and curve-of-growth analyses \citep{schon88,vil02}. Consequently, the status of OBN/OBC stars remains unclear, and the origins of the chemical-abundance peculiarities observed at their surfaces are not fully understood -- e.g., what are the relative roles of rotation and binarity? In order to make progress in the understanding of OBN stars, we have conducted the first quantitative analysis of the helium and CNO abundances of a significant sample of ON stars, using modern non-LTE/line-blanketed atmosphere models. Section~\ref{s_obs} of the paper presents the sample and the observational material; analysis methods are described in Section~\ref{s_mod},  and the results in Section~\ref{s_res}. These results are discussed in Section~\ref{s_disc}, with our conclusions given in Section~\ref{s_conc}.

\section{Observations and sample}
\label{s_obs}

The most complete catalogue of Galactic late-type ON stars known to date is provided by \citet{walborn11}. It contains 13 stars: four dwarfs/sub-giants, seven giants, and two supergiants.\footnote{\citet{walborn11} explicitly exclude two further stars. The first, BD+36~4063, is an O9.7\;Iab supergiant \citet{mathys89}, known to be a spectroscopic binary \citep{williams09}. The second, HD~105056, is classified as ON9.7\;Iae (e.g., \citealt{jesus04}), but it may be a lower-mass, post-AGB object \citep{walborn11}.} In the present study, we consider twelve stars from this sample. We exclude the ON9.7\;II--III(n) star HD~89137 because it is a double-lined spectroscopic binary and the available data are not sufficient to separate unambiguously the two components.

The spectroscopic data were collected from a number of archives and
unpublished material; sources and dates of observation are listed in
Table \ref{tab_obs}. We give a brief description of each dataset 
below:

\begin{itemize}

\item AAT data were acquired at the 3.9-m Anglo-Australian Telescope
with the University College London Echelle Spectrograph (UCLES) and a
31.6 lines mm$^{-1}$ grating.  Observations were made at
several spectrogrograph settings in order to obtain near-continuous
wavelength coverage $\sim$3650--7000\AA\ at a resolving power $R \simeq 23\,000$, and
signal-to-noise levels $\gtrsim$100 throughout.

\item The IACOB spectroscopic database \citep{sergio11} includes multi-epoch observations for seven of the stars listed in Table \ref{tab_obs}. The observations, obtained with the FIbre-fed Echelle Spectrograph \citep[FIES, ][]{telting14} attached to the 2.5~m Nordic Optical Telescope (NOT), cover the spectral range 3700--7300\AA\ with $R$=46\,000. Data reduction was performed with FIEStool\footnote{http://www.not.iac.es/instruments/fies/fiestool/} and the normalization of the spectra using own procedures developed in IDL.

\item Spectra of four targets were extracted from the ESO/FEROS
archive. These data were obtained in the context of the OWN project
\citep{barba10,barba14}. They have a resolving power of 48\,000 and were
reduced by the fully automated pipeline distributed by ESO.

\item HD~191423 was observed with the Elodie Spectrograph, mounted on the 1.93-m telescope at Observatoire de Haute-Provence (OHP). The resolving power is 42\,000 and the wavelength coverage 3895--6815 \AA.  Exposure times were chosen to ensure a signal-to-noise ratio of at least 100 at 5200 \AA. Reduction was performed using the standard reduction pipeline described in \citet{bar96}.

\item For HD~14633 and HD~201345, spectra were acquired with the ESPaDOnS spectropolarimeter mounted on the Canada--France--Hawaii Telescope. The resolving power is 65\,000 and the wavelength coverage is 3700--10\,500 \AA. Data reduction was performed with the automated procedure \textit{Libre Esprit}; a full description of the data is given by Wade et al.\ (2015, submitted).

\end{itemize}

Table~\ref{tab_obs} lists the observational data used for the spectroscopic analysis. Additional data are described in Table~\ref{tab_var} and will be discussed in Sect.~\ref{s_var}.

\begin{table}
\begin{center}
\caption{Observational information.} \label{tab_obs}
\begin{tabular}{lccc}
\hline
Star        & ST        &  Instrument & date of observation\\    
            &           & \\
\hline
HD~12323    & ON9.2 V    & FIES       & 08 sep 2011 \\
HD~13268    & ON9.5 IIIn  & FIES       & average \\
HD~14633    & ON8.5 V    & ESPaDOnS   & 09 oct 2009 \\      
HD~48279    & ON8.5 V    & FIES     & average  \\ 
HD~91651    & ON9.5 IIIn  & FEROS      & 05 may 2009 \\ 
HD~102415   & ON9 IIInn     & FEROS      & 10 jun 2008 \\ 
HD~117490   & ON9.5 IIInn  & FEROS      & 14 may 2008  \\ 
HD~123008   & ON9.2 Iab  & UCLES    & 22 jun 1992  \\ 
HD~150574   & ON9 III(n)    & FEROS      & 10 jun 2008 \\ 
HD~191423   & ON9 II-IIInn & ELODIE     & 29 aug 2004 \\ 
HD~191781   & ON9.7 Iab  & UCLES     & 14 aug 1995\\
HD~201345   & ON9.2 IV   & ESPaDOnS   & 26 jul 2010  \\  
\hline
\end{tabular}
\tablefoot{Spectral types are from \citet{sota11,sota14}.}
\end{center}
\end{table}

\section{Modelling and spectroscopic analysis}
\label{s_mod}

We have used the atmosphere code CMFGEN to analyze the surface properties of the ON stars. A full description of CMFGEN can be found in \citet{hm98}. In a nutshell, CMFGEN solves the radiative transfer and statistical equilibrium equations in the comoving frame, leading to non-LTE models. The temperature structure is set from the radiative-equilibrium constraint. Spherical geometry is adopted to take into account extension due to the strong winds of O stars. The density structure is computed from mass conservation and the velocity structure is constructed from a pseudo-photospheric structure connected to a $\beta$-velocity law. The photospheric structure is obtained from a few iterations of the hydrodynamical solution in which the radiative force computed from the level populations and atomic data is included. The final synthetic spectrum is obtained from a formal solution of the radiative transfer equation. 

The synthetic spectra are subsequently compared to observations to determine the stellar parameters. In practice, we proceeded as follows:

\begin{itemize}

\item \emph{Rotation and macroturbulence:} we used the Fourier-transform method \citep{gray76} to determine \vsini, the projected equatorial rotational velocity. We relied on \ion{O}{iii}~$\lambda$5592 when possible; otherwise, we used \ion{He}{i}~$\lambda$4920. We subsequently estimated the expected \teff\ and \logg\ from the target's spectral type, using the calibration of \citet{msh05}, and selected a corresponding synthetic spectrum form our database of models. We convolved this spectrum with a rotational profile (adopting \vsini\ from the previous step), and performed an additional convolution by a radial--tangential profile (parameterized by a velocity \vmac) to take macroturbulence into account \citep{sergio14}. We varied \vmac\ until a good match was obtained to the observed spectrum (especially around 4100 \AA, 4700 \AA\ and 4900 \AA). Uncertainties on \vsini\ and \vmac\ are of the order $\sim$10 and 20 \kms\ respectively. 

\item \emph{Effective temperature:} we relied on the traditional ionization-balance method to constrain \teff. We used the helium lines \ion{He}{i}~$\lambda$4026, \ion{He}{ii}~$\lambda$4200, \ion{He}{i}~$\lambda$4388, \ion{He}{i}~$\lambda$4471, \ion{He}{ii}~$\lambda$4542, \ion{He}{i}~$\lambda$4712, \ion{He}{i}~$\lambda$4920, \ion{He}{i}~$\lambda$5016, and \ion{He}{ii}~$\lambda$5412. We found that when \ion{He}{ii}~$\lambda$4542 was perfectly matched, \ion{He}{ii}~$\lambda$4200 was usually slightly too strong and  \ion{He}{ii}~$\lambda$5412 slightly too weak. This is partly attributed to the échelle nature of most of our spectra and thus to uncertain normalization. As a consequence, effective temperatures are determined within about $\pm$1500~K. 

\item \emph{Surface gravity:} the wings of Balmer lines were used to determine \logg, with a typical uncertainty of $\sim$0.15~dex.

\end{itemize}

Since our principal focus is on surface parameters, we did not try to reproduce the details of emission lines likely to be formed in the wind (particularly \ion{He}{ii}~$\lambda$4686, H$\alpha$);  instead, we simply adopted combinations of terminal velocity and mass-loss rate that give a reasonable fit to these features in each star, with the wind-acceleration parameter $\beta$ fixed at 1.0. Luminosities were adopted from the calibration given by \citet{msh05}. Instrumental broadening was negligible given the high resolution of the observed spectra and the generally large rotational broadening, and our final synthetic spectra were simply convolved by rotational and radial-tangential profiles (see above).

\begin{figure}[]
\centering
\includegraphics[width=9cm]{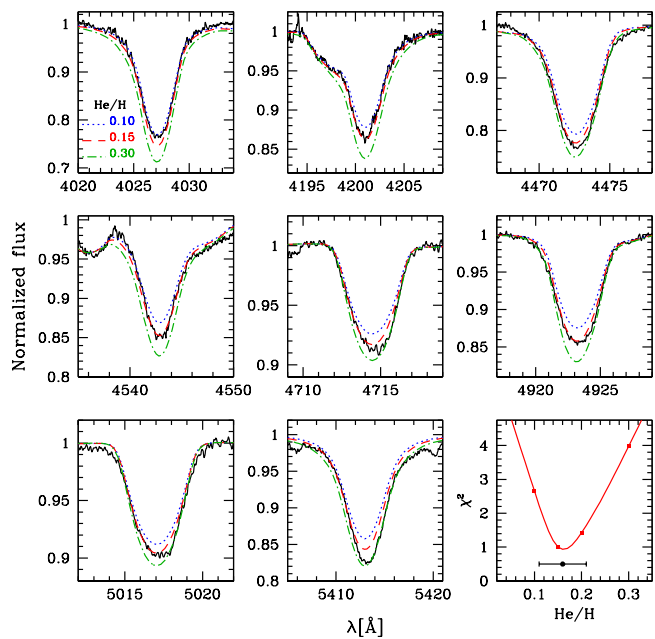}
\caption{Determination of the ratio He/H for HD~48279. Observed \ion{He}{i} and \ion{He}{ii} lines are shown by the black solid line. The blue dotted (red dashed; green dot-dashed) lines are models with He/H=0.10 (0.15; 0.30). The lower right panel shows the $\chi^{2}$ results from the quantitative analysis, from which one determines He/H=0.16$\pm$0.05.}
\label{fig_heh}
\end{figure}

For the determination of surface abundances, we proceeded as in \citet{mimesO}. For a given \teff\ and \logg\ we ran models with different He, C, N, and O abundances. We identified the cleanest lines of each element in the observed spectrum, and performed a $\chi^2$ analysis combining all selected features. The computed $\chi^2$ values were renormalized to a minimum value of 1.0, with the best-fit abundance taken to correspond to this minimum, and 1-$\sigma$ uncertainties defined by $\chi^2$=2.0.

For helium, we utilized from seven to ten lines selected from \ion{He}{i}~$\lambda$4026, \ion{He}{ii}~$\lambda$4200, \ion{He}{i}~$\lambda$4388, \ion{He}{i}~$\lambda$4471, \ion{He}{ii}~$\lambda$4542, \ion{He}{i}~$\lambda$4713, \ion{He}{i}~$\lambda$4920, \ion{He}{i}~$\lambda$5016, \ion{He}{ii}~$\lambda$5412, and \ion{He}{i}~$\lambda$6680. Figure \ref{fig_heh} shows an example of He/H determination, illustrated by HD~48279. An abundance greater than 0.1 (by number) is clearly favoured, simply by inspection. The quantitative analysis restablishes that He/H=0.16$\pm$0.05  best reproduces the set of helium lines.  

\begin{figure}[]
\centering
\includegraphics[width=9cm]{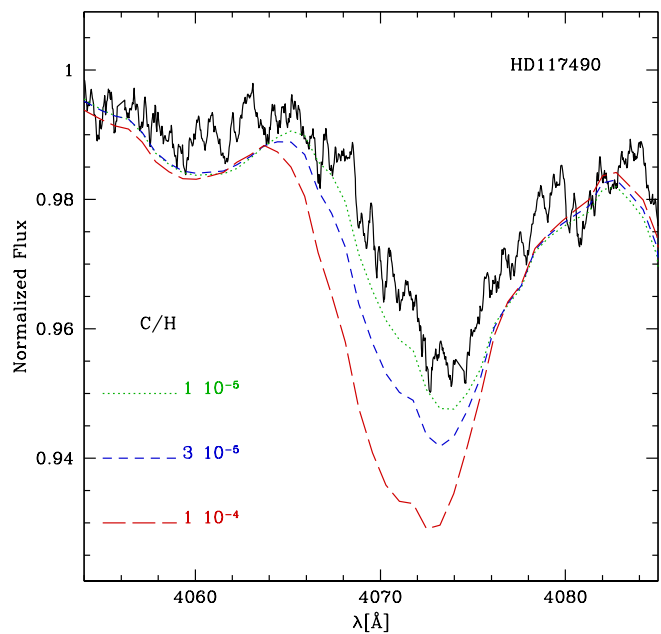}
\caption{Observed \ion{C}{iii}~4070 line of HD~117490 (black solid line) together with models with different C/H values (colored interrupted lines). }
\label{fig_conh}
\end{figure}

Our primary carbon-abundance diagnostic was \ion{C}{iii}~$\lambda$4070. For a few stars, \ion{C}{iii}~$\lambda$4163, \ion{C}{iii}~$\lambda$4188, and \ion{C}{ii}~$\lambda$4267 could also be used. We did not use \ion{C}{iii}~$\lambda$4650 because of its sensitivity to winds, to metallicity and to uncertainties in the atomic data \citep{mh12}. In most cases, we could obtain only an upper limit on C/H  because \ion{C}{iii}~$\lambda$4070 is the only useful carbon line that could be detected, and is rather weak. Difficulties in normalizing the spectra resulting from the presence of nearby H$\delta$ further undermine any attempt to put a better constraint on C/H. Figure \ref{fig_conh} shows an example of C/H determination, for HD~117490. In this case a conservative upper limit on C/H of $3 \times\ 10^{-5}$ is adopted. 

The principal nitrogen-abundance indicator was the \ion{N}{iii} 4510--4535 \AA\ complex of lines. Occasionally, other indicators were added: \ion{N}{ii}~$\lambda$3995, \ion{N}{ii}~$\lambda$4004, \ion{N}{iii}~$\lambda$4044, \ion{N}{ii}~$\lambda$4447, \ion{N}{ii}~$\lambda$4607, \ion{N}{ii}~$\lambda$5001, \ion{N}{ii}~$\lambda$5005, \ion{N}{ii}~$\lambda$5011, \ion{N}{ii}~$\lambda$5676 and \ion{N}{ii}~$\lambda$5680. Finally, \ion{O}{iii}~$\lambda$5592 was adopted as the main oxygen-abundance indicator. In some stars, we could also make use of \ion{O}{ii}~$\lambda$3913, \ion{O}{ii}~$\lambda$3963, \ion{O}{ii}~$\lambda$4277-78, \ion{O}{ii}~$\lambda$4318, \ion{O}{ii}~$\lambda$4368, \ion{O}{ii}~$\lambda$4416-18, \ion{O}{ii}~$\lambda$4603, and \ion{O}{ii}~$\lambda$4611. Compared to normal O stars, fewer useful lines are generally available to determine surface abundances because the ON stars typically have relatively large \vsini\ values \citep{hs01}; hence weak metallic lines are too broad to be readily detected.
Lines from two ionization states are usually taken into account in the abundance determination, so that the error bars include uncertainties related to flawed N and O ionization balances. At the same time, by performing the abundance determination on an ensemble of lines we minimize uncertainties related to the formation/physics of individual lines \citep[see Fig.\ 1 in][]{mimesO}.

We adopted a photospheric microturbulent velocity of 10 \kms\ in our synthetic spectra\footnote{Except for HD~123008 for which we found that 20 \kms\ gave a better fit of the helium lines}. A larger value would tend to reduce the derived abundances, especially of helium \citep{mcerl98,vil00,hs01}. Using HD~191423 as a test case, we find that increasing the microturbulence to 20 \kms\ leads to a reduction of 0.05 in the inferred He/H ratio.

As a check on our results, we performed a separate analysis of our
sample using the {\sc iacob-gbat} package \citep{sergio11,sabin14},
which uses a $\chi^2$-fitting algorithm coupled to a large,
pre-computed grid of FASTWIND models \citep{santo97,puls05}.  This
completely independent approach allows the determination of \teff,
\logg, \vsini, \vmac, and He/H.  The results turned out to be in
excellent agreement with the CMFGEN analyses, within the respective
error bars, lending strong support to our quantitative results.

\section{Results}
\label{s_res}

The results of the spectroscopic analysis are summarized in Table
\ref{tab_param}, and the best-fit models compared to the observed
spectra in Figs.~\ref{fit_12323}--\ref{fit_201345}. We emphasize that
the numerical results listed in Table~\ref{tab_param} should be viewed
as `surface average' values; many of the sample stars have large
rotational velocities, which can affect their shapes, and create
surface gradients in the physical parameters
\citep[e.g.][]{hs01,palate13}.  For HD~191423, the fastest rotator of
the sample, rotational broadening is so large that normalization of
the echelle spectra was extremely difficult, and carbon and oxygen
abundances could not be determined.

\begin{table*}
\begin{center}
\caption{Parameters of the sample stars.} \label{tab_param}
\begin{tabular}{lcccccccccc}
\hline
Star        & Spectral     &   Teff  &   logg & logg$_c$&  \vsini\  &  \vmac  &  He/H  &  C/H         &    N/H      & O/H   \\    
            &  Type        & [kK]    &        &         &  [\kms]   &  [\kms] &        &  [10$^{-4}$]  &  [10$^{-4}$] &  [10$^{-4}$] \\
\hline
HD~12323    & ON9.2 V      & 33.5 & 4.00  & 4.01  &  130 & -   & 0.16$^{+0.10}_{-0.06}$ & $<$0.3 & 3.5$^{+2.7}_{-1.8}$ & 2.5$^{+2.3}_{-1.6}$ \\ 
HD~13268    & ON9.5 IIIn   & 32   & 3.50  & 3.63  &  310 & -   & 0.20$\pm$0.10 &  $<$0.5     &  5.0$^{+2.8}_{-2.0}$ & 3.1$^{+2.5}_{-1.4}$ \\
HD~14633    & ON8.5 V      & 34   & 3.80  & 3.81  &  100 & 120 & 0.13$^{+0.08}_{-0.04}$ &  $<$0.3     &  5.2$^{+5.0}_{-2.5}$  & 1.7$^{+0.9}_{-0.8}$ \\ 
HD~48279    & ON8.5 V      & 34.5 & 3.80  & 3.82  &  137 & 50  & 0.16$\pm$0.05 & 0.65$^{+0.20}_{-0.20}$ & 4.6$^{+3.2}_{-2.0}$ & 4.3$^{+2.0}_{-2.3}$ \\
HD~91651    & ON9.5 IIIn   & 31   & 3.50  & 3.62  &  310 & -   & 0.14$^{+0.08}_{-0.05}$ & $<$0.5 & 5.4$^{+1.6}_{-1.1}$ & 2.3$^{+1.9}_{-1.4}$ \\
HD~102415   & ON9 IIInn    & 31   & 3.50  & 3.70  &  376 & -   & 0.21$\pm$0.10 &  $<$0.6    &  7.6$^{+4.2}_{-3.6}$  & $<$3.0 \\
HD~117490   & ON9.5 IIInn  & 30.5 & 3.50  & 3.66  &  375 & -   & 0.16$^{+0.09}_{-0.05}$ & $<$0.3  &  7.6$^{+2.9}_{-2.6}$ & 2.5$^{+2.6}_{-2.0}$\\
HD~123008   & ON9.2 Iab    & 30   & 3.10  & 3.10  &  37  & 100 & 0.21$^{+0.11}_{-0.07}$ & 0.22$^{+0.21}_{-0.18}$ & 13.5$^{+6.7}_{-3.6}$ & 4.8$^{+2.1}_{-2.1}$ \\
HD~150574   & ON9 III(n)   & 31   & 3.40  & 3.49  &  240 & -   & 0.23$\pm$0.06 &  $<$0.5 & $>$10.0 & 6.0$^{+2.9}_{-2.3}$ \\
HD~191423   & ON9 II-IIInn & 31.5 & 3.50  & 3.72  &  445 & -   & 0.25$\pm$0.08 &  --   &  $>$5.0  & -- \\
HD~191781   & ON9.7 Iab    & 28   & 3.10  & 3.12  &  107 & 70  & 0.25$^{+0.2}_{-0.1}$ & $<$0.5 &  7.3$^{+7.3}_{-5.0}$ & 3.1$^{+0.8}_{-0.8}$ \\
HD~201345   & ON9.2 IV     & 34   & 4.00  & 4.01  &  95  & 60  & 0.1          &  $<$0.4     &  4.0$^{+2.2}_{-1.3}$   & 3.6$^{+2.7}_{-2.2}$ \\
\hline
\end{tabular}
\tablefoot{Uncertainties on \teff, \logg, \vsini\ and \vmac\ are $\sim$1.5kK, 0.15 dex, 10 and 20 \kms\ respectively. logg$_c$ is the surface gravity corrected for centrifugal acceleration. Abundances are number ratios.}
\end{center}
\end{table*}

Table \ref{tab_param} shows that most ON stars are helium-rich. \citet{sh94} performed a helium-abundance determination for HD~123008 and obtained He/H$=0.20\pm 0.05$, in good agreement with our estimate. \citet{hs01} determined He/H in HD~191423 using simple non-LTE models without line-blanketing but a better treatment of 2D effects than our approach. They obtained He/H$=0.23\pm 0.04$ in very good agreement with our value. The relatively large helium enrichment of most ON stars indicates a peculiar chemical history compared to normal OB stars, which do not show systematically values of He/H larger than 0.1 \citep[e.g.][]{mokiem07}.

\begin{figure}[t]
\centering
\includegraphics[width=9cm]{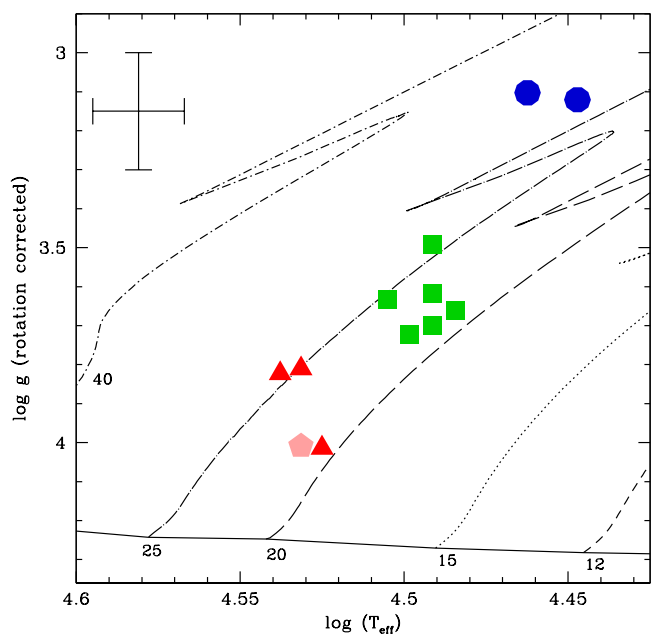}
\caption{\logg\ - log(\teff) diagram for the sample stars. Triangles (pentagons, squares, circles) are for luminosity class V (IV, III, I) stars. Typical uncertainties are shown in the upper-left corner. Evolutionary tracks including rotation, from \citet{ek12}, are overplotted, labelled by ZAMS masses.}
\label{fig_hr}
\end{figure}

Figure \ref{fig_hr} shows the surface gravity as a function of effective temperature in our sample. 'Geneva' evolutionary tracks, from \citet{ek12}, are overplotted. We see that the ON stars follow a relatively clear sequence: dwarfs have higher \logg\ values than giants, while supergiants have the lowest \logg, a trend similar to that found for normal O stars by \citet{mimesO}. Stars classified as ON have a rather narrow range of initial masses: 20--25~\msun\ for the dwarfs/giants, and a little over 25~\msun\ for the supergiants, according to the Geneva tracks.

\begin{figure}[t]
\centering
\includegraphics[width=9cm]{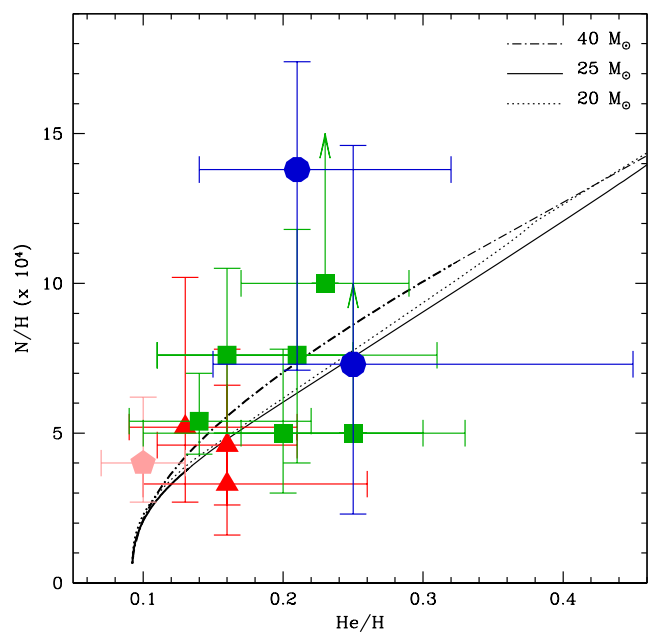}
\caption{N/H as a function of He/H for the ON stars. The abundance ratios are by number. Symbols have the same meaning as in Fig.~\ref{fig_hr}. The evolutionary tracks including rotation of \citet{ek12} are overplotted. The bold part corresponds to the main sequence (central hydrogen mass fraction $>$ 0).}
\label{n_he}
\end{figure}

Figure~\ref{n_he} illustrates the N/H ratio (by number) as a function
of He/H. The evolutionary models predict a correlation between N/H and
He/H that simply results from nucleo{\-}synthesis through the CNO
cycle: the higher the helium content, the higher the nitrogen
abundance. Figure~\ref{n_he} shows a possible trend of this kind among
the ON stars, but the rather large error bars on the abundance
determinations don't allow for a clear-cut conclusion
observationally. Most stars appear more enriched than expected from
their position in the \logg\ - log(\teff) diagram.

\begin{figure}[t]
\centering
\includegraphics[width=9cm]{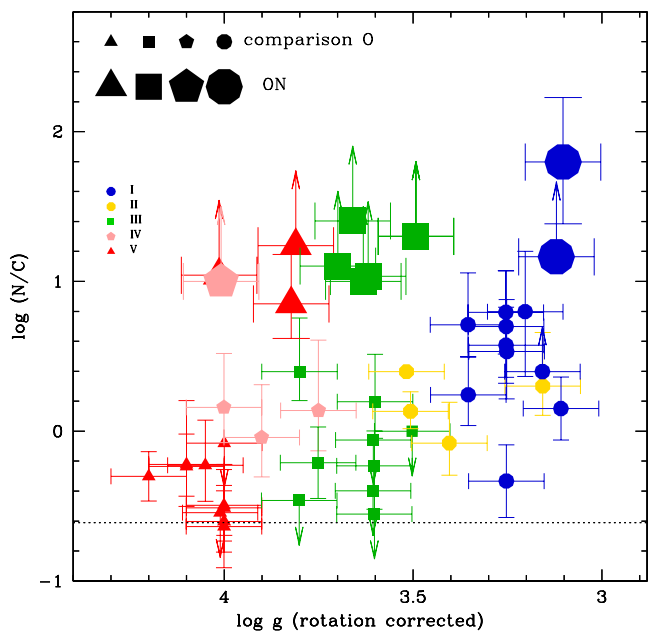}
\caption{The N/C abundance ratio as a function of surface gravity for the ON stars (large symbols), compared to results for morphologically normal stars with spectral types O8.5--9.7 (small symbols), taken from \citet{mimesO}.}
\label{nonc_logg}
\end{figure}

Figure \ref{nonc_logg} shows the ratio of nitrogen to carbon surface
abundance as a function of surface gravity for the ON
stars\footnote{As discussed previously, only upper limits to C/H are
  available in most cases.} and for a reference sample of
morphologically normal O stars taken from \citet{mimesO}. For the
reference sample, only stars with spectral types between O8.5 and O9.7
are considered, corresponding to the spectral-type range of the ON
sample. The dwarf, giant, and supergiant stars are well separated in
this plane, among both ON and reference groups, largely reflecting the
surface-gravity differences between luminosity classes, as is also
seen in Fig.~\ref{fig_hr}.  The most striking feature of
Fig.\ \ref{nonc_logg} is the higher log(N/C) ratios found for ON
stars, for a given luminosity class. The difference amounts to 0.5-1.0
dex or more, since most of the log(N/C) values for ON stars are
actually lower limits (due to the upper limits on C/H; Table
\ref{tab_param}). Another possible trend is that of higher log(N/C)
among ON stars as one moves from dwarfs to giants and
supergiants. This trend was clearly established for morphologically
normal stars by \citet{mimesO}, and appears to hold also for ON
stars. This suggests that the physics of chemical mixing follows the
same patterns in normal and ON stars, pointing to a common origin for
surface chemical enrichment. However, we caution that the trend for ON
stars is less clear due to both the small number of objects, and the
fact that for most stars we have only lower limits on log(N/C).

\begin{figure*}[t]
\centering
\includegraphics[width=0.45\textwidth]{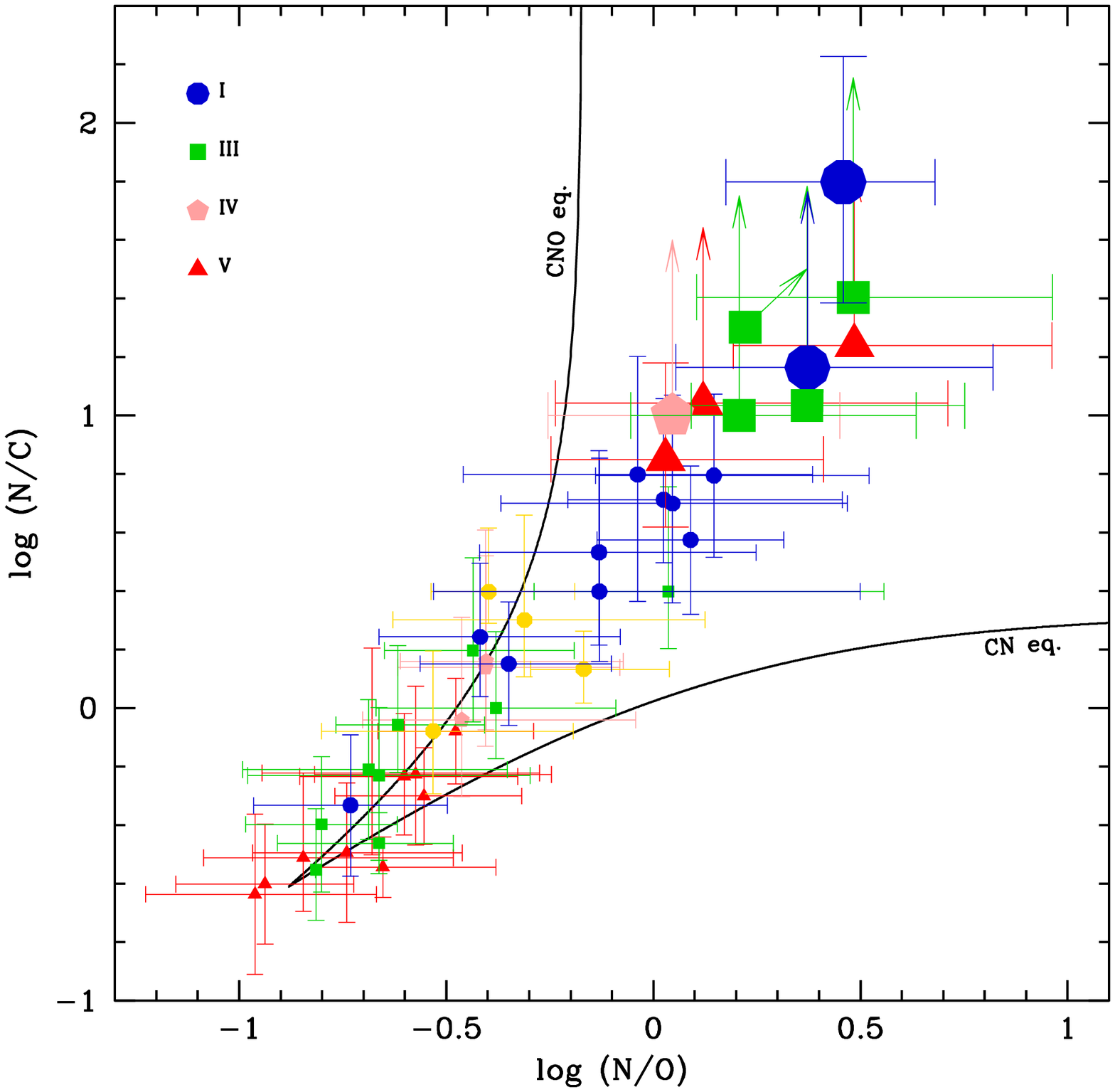}
\includegraphics[width=0.45\textwidth]{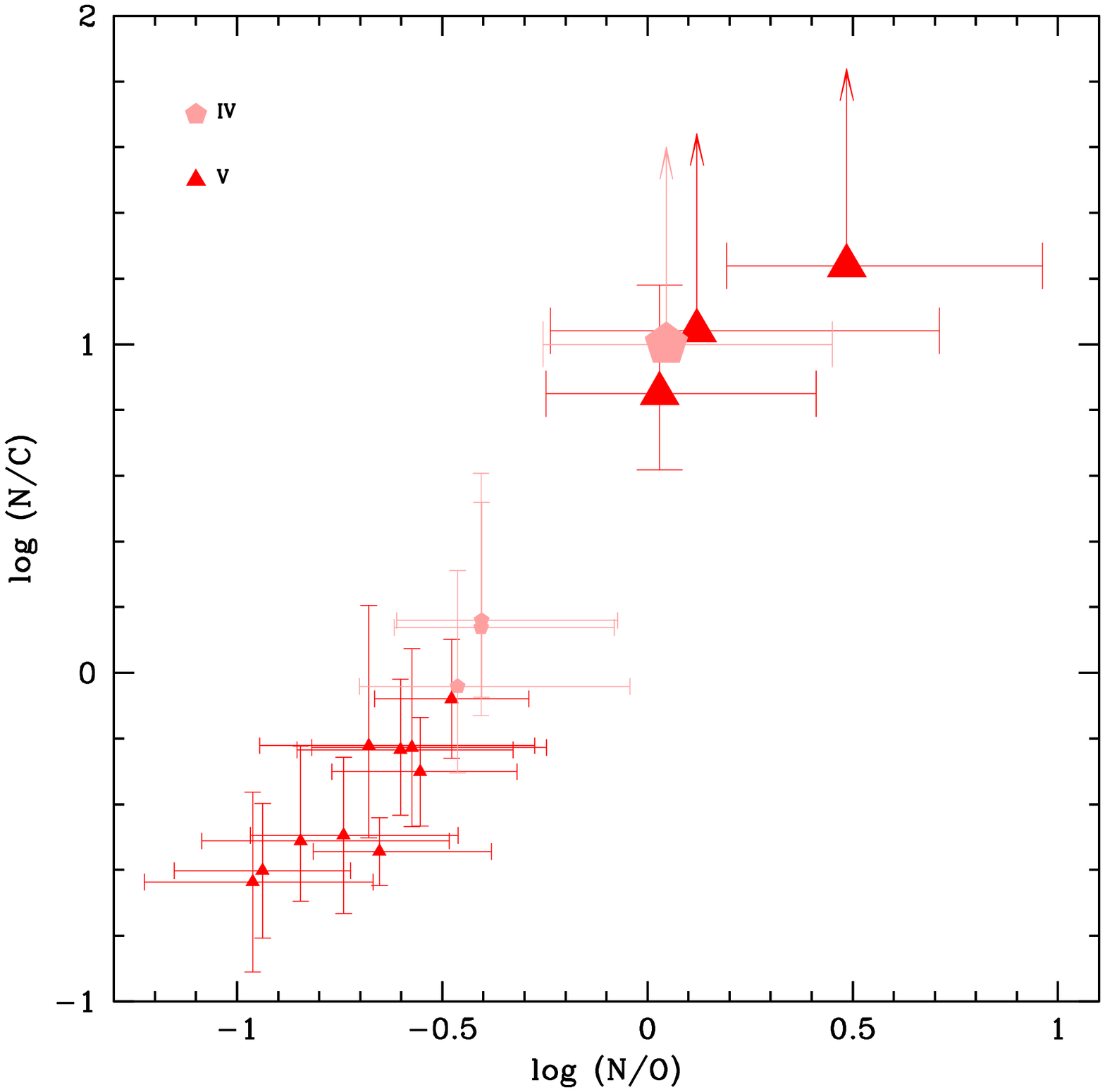}\\
\includegraphics[width=0.45\textwidth]{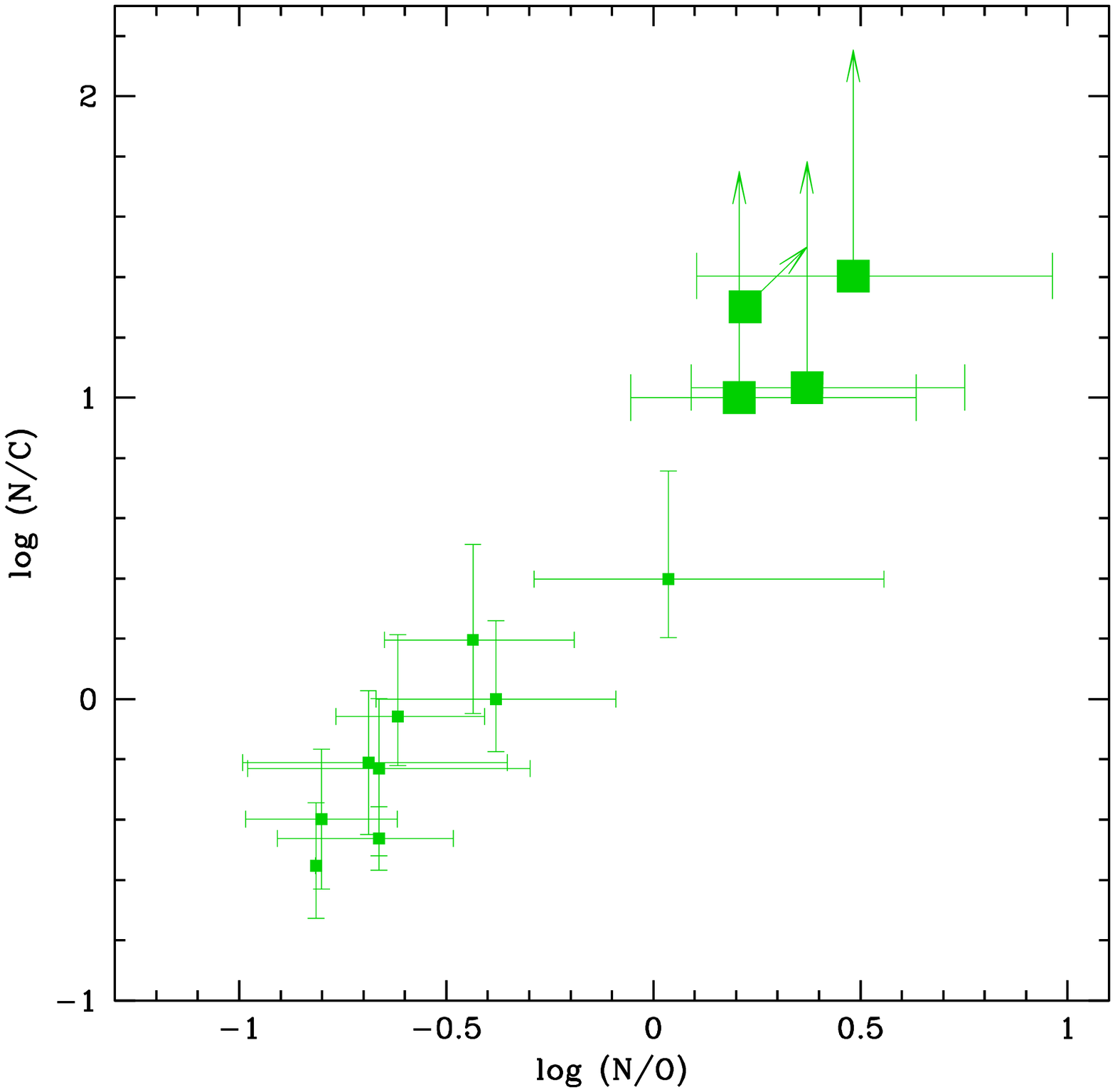}
\includegraphics[width=0.45\textwidth]{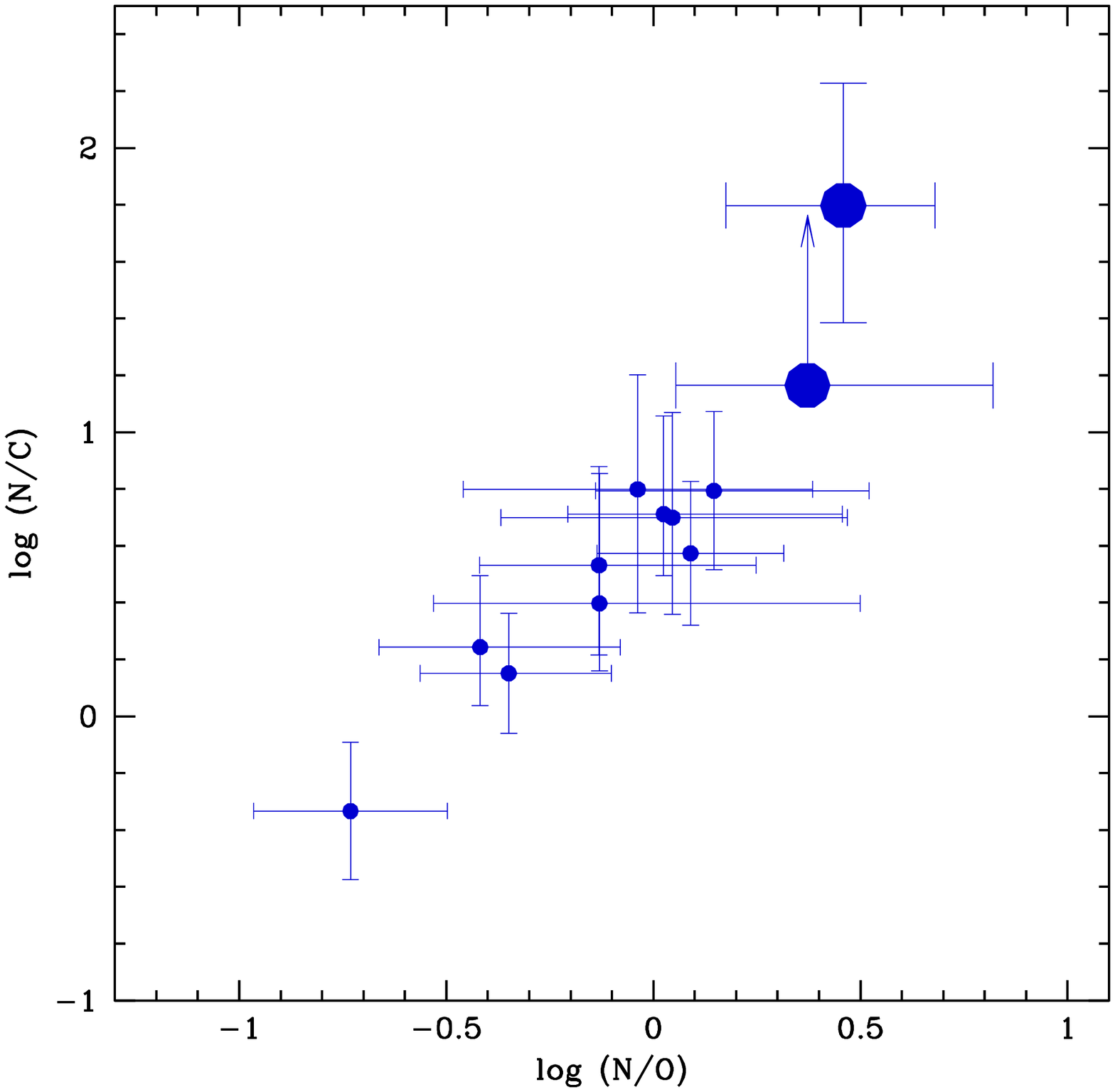}
\caption{Distributions of log (N/C) versus log(N/O) for the ON stars (large symbols) and for a comparison set of normal O stars with spectral types O8.5-O9.7 (small symbols) taken from \citet{mimesO}. The upper-left panel shows the full sample. The upper-right (lower-left, lower-right) shows the dwarf--subgiant (giant, supergiant) sample. Solid lines in the upper-left panel show the relations expected for partial CN or complete CNO burning in equilibrium.}
\label{fig_cno}
\end{figure*}

Figure~\ref{fig_cno} shows the log(N/C) versus log(N/O) diagram. The
upper-left panel reveals that the ON stars are systematically more
enriched than the comparison O stars taken from \citet{mimesO}. They
extend the relation between log(N/C) and log(N/O) observed for normal
stars to higher abundance ratios, and are also located between the
limits corresponding to partial CN burning and complete CNO
burning. Consequently, the abundance patterns of ON stars are
consistent with the predictions of nucleo{\-}synthesis through the CNO
cycle.

Other panels of Fig.~\ref{fig_cno} compare results for ON and normal O
stars at different luminosity classes. In each case, the ON stars are
clearly separated from the comparison stars, indicating that their
surface abundances result from a much stronger mixing than that
experienced by normal stars. \citet{mimesO} showed that, on average,
supergiants are more enriched than giants and dwarfs. We saw in
Fig.~\ref{nonc_logg} that this trend may exist among ON stars,
too. Fig.~\ref{fig_cno} tends to confirm that there is a sequence of
higher enrichment when moving from dwarfs to giants and supergiants,
although there is one outlier (HD~14633, a dwarf that appears to be as
chemically mixed as the giants).

\begin{figure}[t]
\centering
\includegraphics[width=9cm]{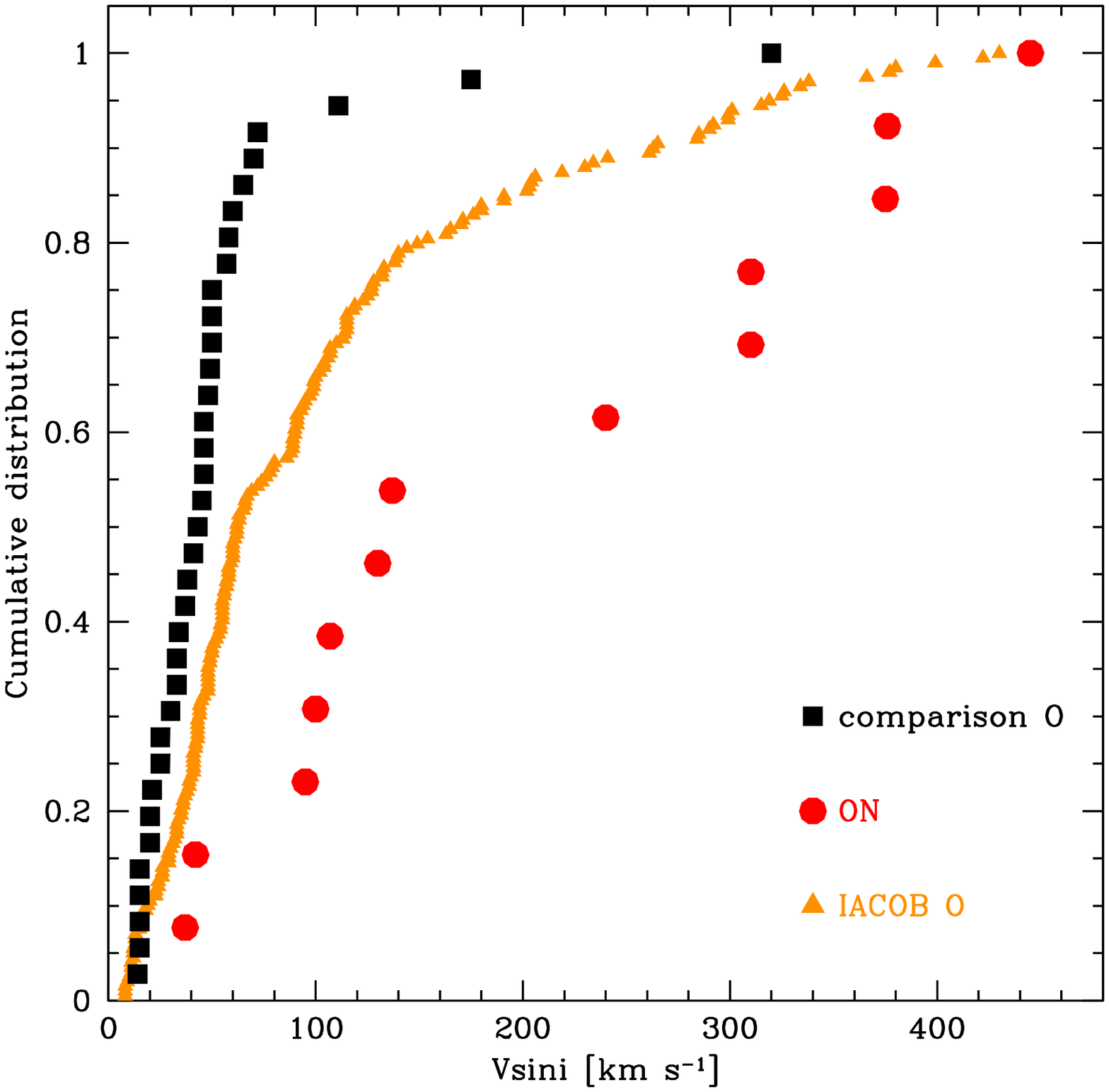}
\caption{Cumulative distribution functions of projected rotational velocities for the ON sample (red circles), our comparison sample (black squares), and the IACOB O-star sample (\citealt{sergio14}; orange triangles).}
\label{vsini_dist}
\end{figure}

Figures~\ref{nonc_logg} and \ref{fig_cno} show gaps between the
distributions of the ON stars and those of the comparison stars, and
the reader may wonder whether these are real features.  In this
context, then supposing that chemical mixing may be due to rotation
(and \citealt{hs01} have shown that ON stars rotate on average faster
than normal O stars), it is informative to consider the distributions
of projected rotational velocities in the two samples.  This
comparison is performed in Fig.~\ref{vsini_dist}; clearly, on average
the stars in the comparison sample are rotating much more slowly than
those of the ON stars. The median \vsini\ value for the ON and
comparison-O samples are 208.0 and 52.5~\kms, and a KS test indicates
a probability of less than 1\% that the two populations are drawn from
the same parent distribution.

Fig.~\ref{vsini_dist} further includes the \vsini\ distribution of the
IACOB sample studied by \citet{sergio14}. They provided projected
rotational velocities for 199 Galactic O stars, which can be viewed as
a reference distribution for O stars in general; the ON stars have
higher \vsini\ values, and our comparison O stars lower ones, than the
\citeauthor{sergio14} reference sample.  Potentially, therefore, the
very different ranges of \vsini\ covered by our two samples (ON and
comparison) can explain the gap seen in Fig.\ \ref{fig_cno}: if
rotation leads to a continuous increase of mixing as
\vsini\ increases, and if our two samples are well separated in terms
of projected rotational velocity, then a separation into two groups in
the log(N/C)--log(N/O) and log(N/C)--\logg\ diagrams is expected.

\section{Discussion}
\label{s_disc}

\subsection{Spectral classification and ON stars}

The spectroscopic classification qualifier `N' was developed in the
context of late-O/early-B stars, as is illustrated by our sample,
which is limited to spectral types between O8.5 and O9.7.
\citet{walborn04} discovered a related morphological dichotomy in
terms of CNO features at the earliest spectral type, O2, where some
stars show relatively strong \ion{O}{iv} lines and weak \ion{N}{iv}
lines. The opposite behaviour is found in a second group, for which
the spectral classification ON2 was therefore created, by analogy to
the behaviour in late-O stars.

It is important to note that different lines are used when classifying
ON2 and late-type ON stars; lines from more highly ionized elements
are observed in the former group (\ion{O}{iv}, \ion{N}{iv}) than in
the latter (\ion{C}{iii}, \ion{N}{iii}). This, rather than some
fundamental difference in surface chemistry, largely accounts for why
the `N' qualifier is not assigned to O stars at intermediate spectral
types -- when moving from O2 to later spectral types, the diagnostic
\ion{O}{iv} and \ion{N}{iv} lines disappear because of ionization
effects.On the other side of the classification scheme, the
\ion{C}{iii} and \ion{N}{iii} lines around 4630--4650\AA, which define
the late-type ON category, progressively go into emission at spectral
types earlier than O8. In principle, the strength of these emission
features could be used in a similar manner to their absorption
counterparts in late-type stars, but the strength of the emission in
these lines is complicated by wind effects \citep{rivero11,mh12}.
Interestingly, \citet{walborn10} recently defined the Ofc class,
corresponding to stars showing \ion{N}{iii}~$\lambda$4630-4640 in
emission and \ion{C}{iii}~$\lambda$4650 of comparable strength. The
difference between Ofc and Of stars may be the equivalent of the ON
phenomenon at O2 and late-O spectral types (Of stars having
comparatively weaker C lines than Ofc stars).

\subsection{Rotation and the ON phenomenon}

Fast rotation could explain the strong chemical enrichments that we
have found in ON stars, which rotate on average faster than normal O
stars (\citealt{hs01}, and section \ref{s_res}).\footnote{We recall
  that such fast rotation could result from formation processes, or
  may arise through tidal interactions in binary systems \citep[which
    would not necessarily have experienced mass
    transfer;][]{langer08}.}  Theoretical predictions indicate that
faster rotation leads to stronger mixing
\citep{mm00,langer12,brott11a,ek12,cl13,ge13}, and thus to the
appearance of more-strongly processed material at the surface of OB
stars.  \citet{walborn04} performed surface-abundance determinations
for O2 and ON2 stars and concluded that the latter were more
chemically processed than the former. From their positions in the
Hertzsprung--Russell diagram, they argued that ON2 stars may be the
product of homogeneous evolution, which is usually understood as a
consequence of fast rotation \citep{maeder87,langer92}.

\begin{figure*}[]
\centering
\includegraphics[width=\textwidth]{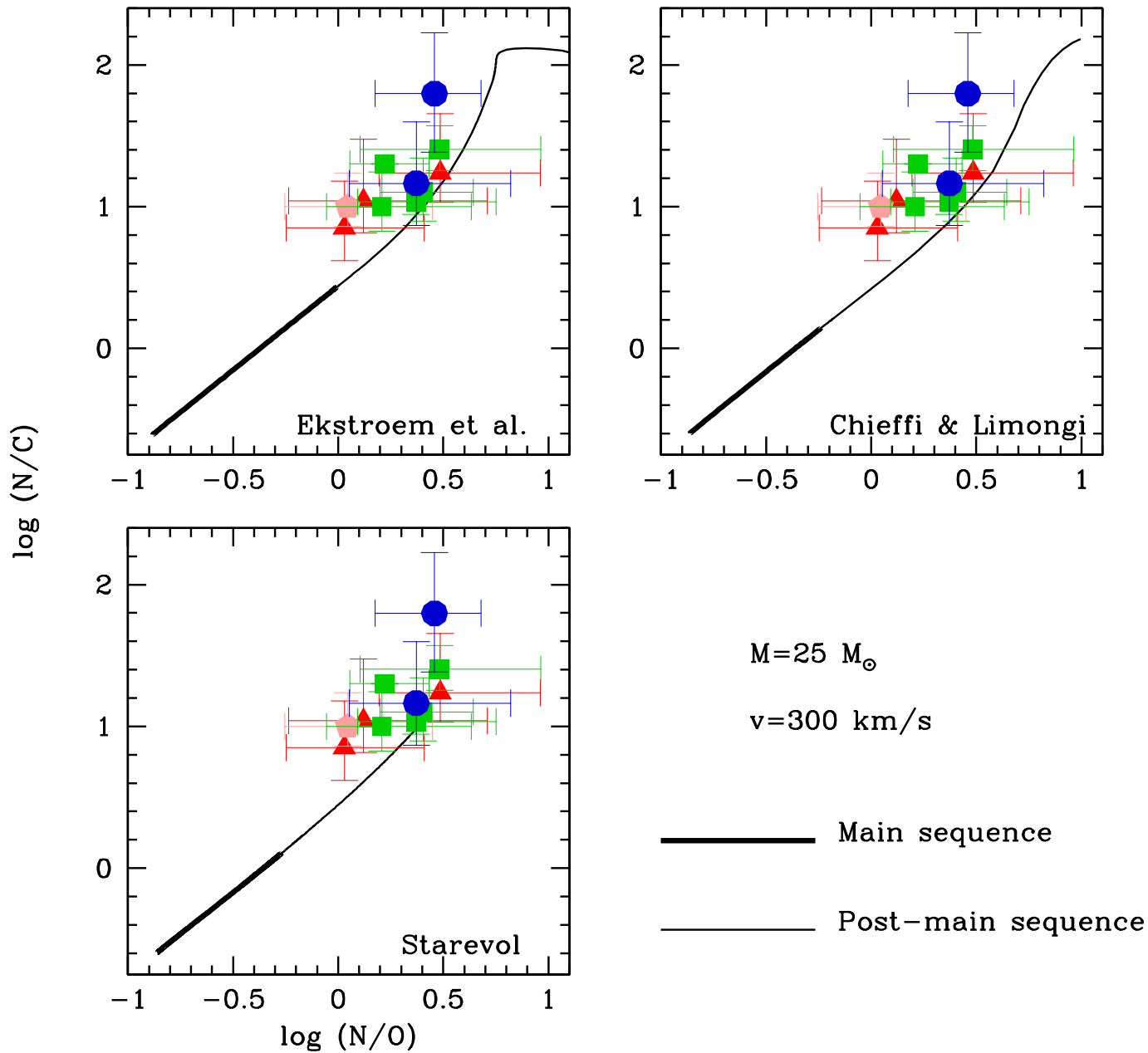}
\caption{As Fig.~\ref{fig_cno}, with evolutionary tracks for  a ZAMS mass of 25~\msun, including rotation (with an initial equatorial velocity of about 300 \kms). Models are from \citet{ek12} (top left), \citet{cl13} (top right), and the STAREVOL code (bottom left). Thick line sections correspond to core hydrogen burning.}
\label{cno_tracks}
\end{figure*}

Figure~\ref{cno_tracks} confronts our abundance determinations with
several stellar-evolution models, computed for 25 M$_\odot$ stars at
solar metallicity, from \citet{ek12}, \citet{cl13}, and a dedicated
model computed with the STAREVOL code \citep{siess06,dmp09} for which
we implemented the same physical ingredients as in \citet{ek12}: core
overshooting, mass-loss prescriptions, angular-momentum-transport
equation, and turbulent-transport prescriptions. All the models start
on the zero-age main sequence with an equatorial surface velocity of
300~\kms\ (equivalent to $V \approx 0.4 V_{\rm c}$, where
$V_{\rm c}$ the critical surface velocity).

All the inferred abundances lie close to the theoretical tracks, which
confirms that they broadly follow the predictions of
nucleo{\-}synthesis and chemical mixing\footnote{The STAREVOL
  calculations were stopped just after the end of the main sequence.}.
However, the abundances predicted at the end of the main sequence
(i.e., when the hydrogen mass fraction in the core reaches zero) vary
significantly between codes, due to the sensitivity of the mixing
efficiency on both the adopted prescriptions for the shear turbulence
(vertical shear from \citet{Maeder97} in \citet{ek12}, and from
\citet{TZ97} in \citet{cl13}), and on the numerical treatment itself.
Nevertheless, we see that the ON stars are \emph{all} located beyond
the end of the main sequence in the log(N/C)--log(N/O) plane.  While
this may be possible for supergiants, the giants and dwarfs are
certainly expected still to be core hydrogen burning (see
Fig.~\ref{fig_hr}).  The surface abundances of ON stars are thus more
enriched than predicted by any of the available models, assuming a
standard rotation rate on the main sequence.

\begin{figure*}[]
\centering
\includegraphics[width=0.32\textwidth]{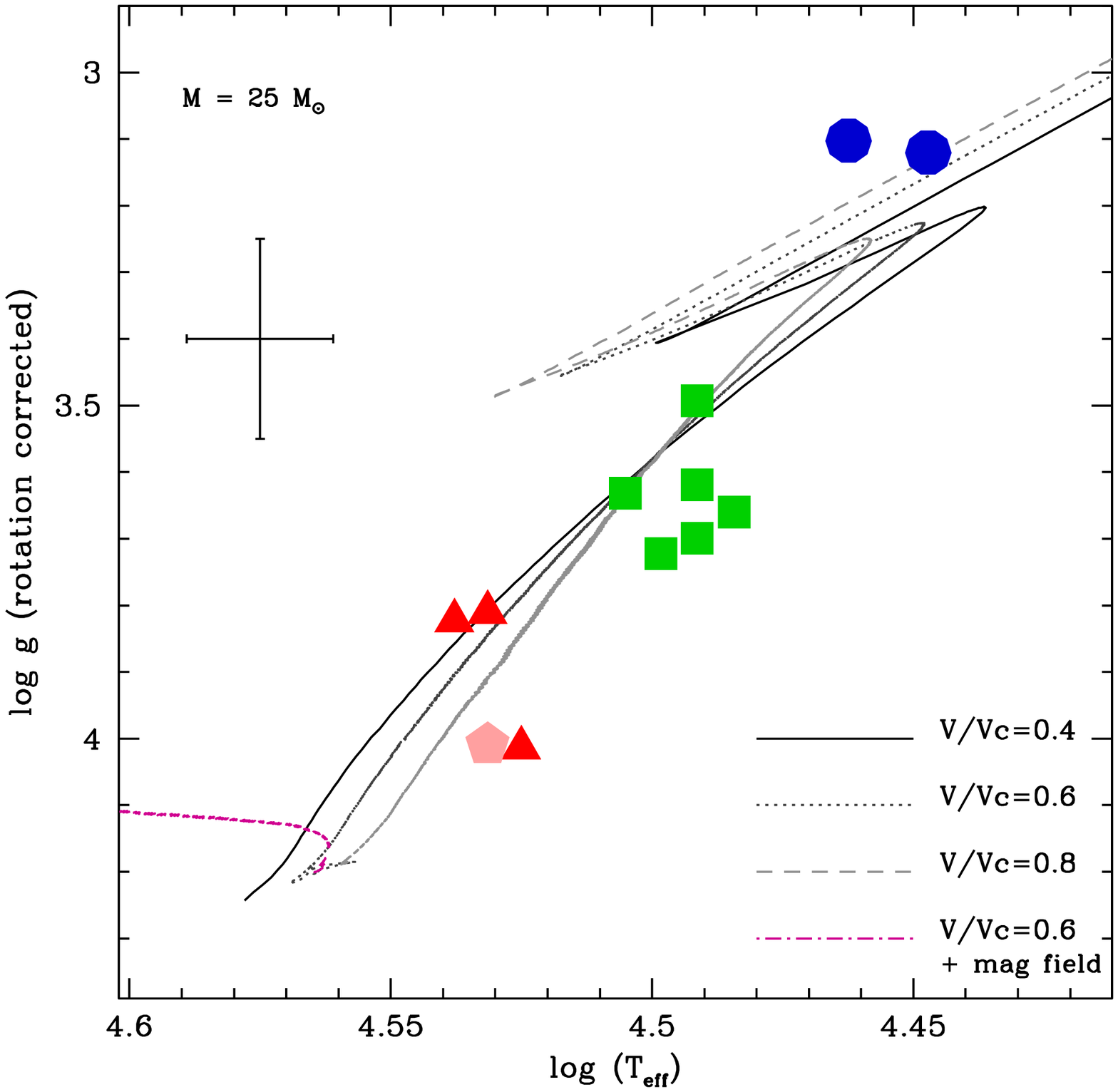}
\includegraphics[width=0.32\textwidth]{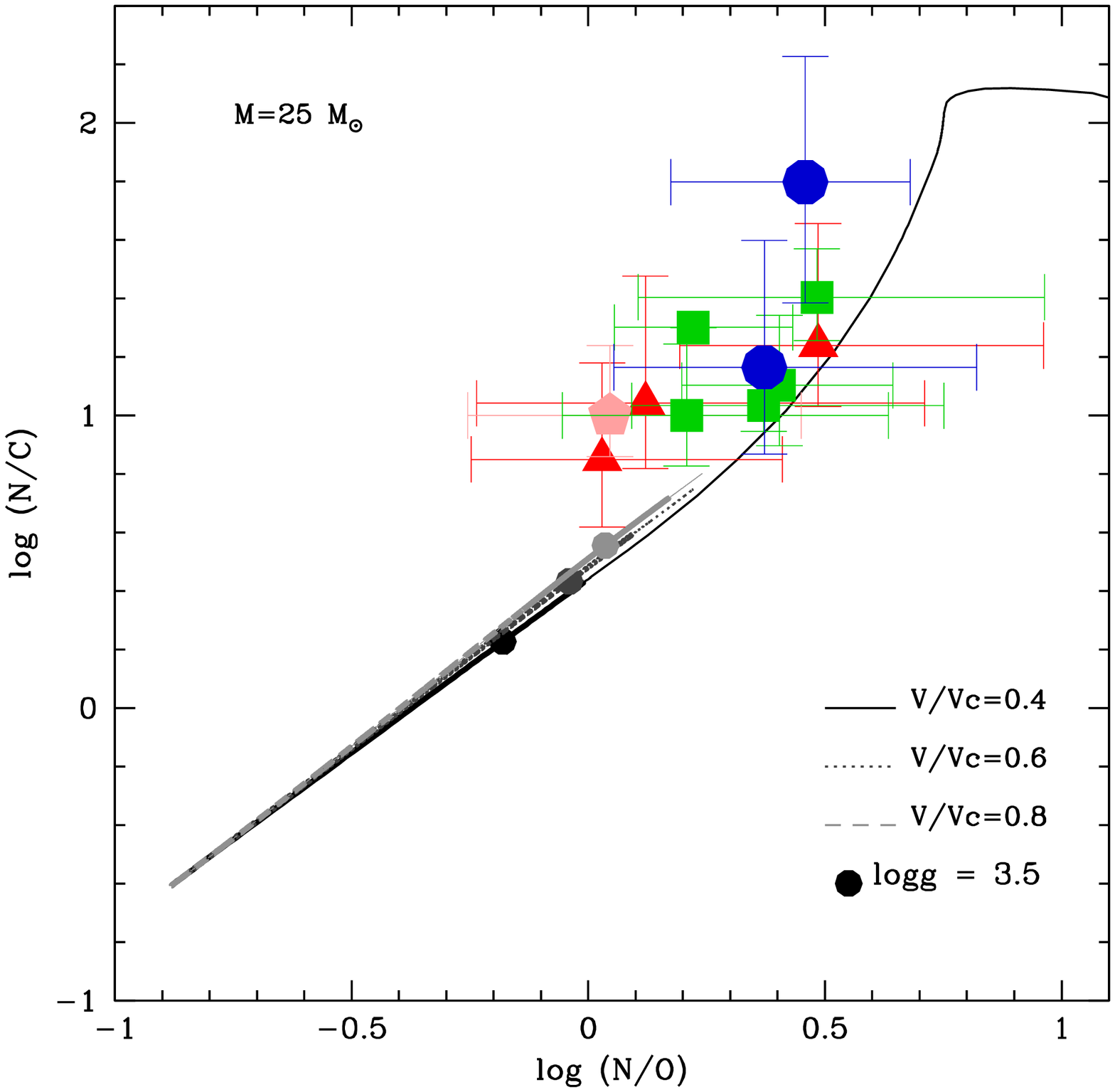}
\includegraphics[width=0.32\textwidth]{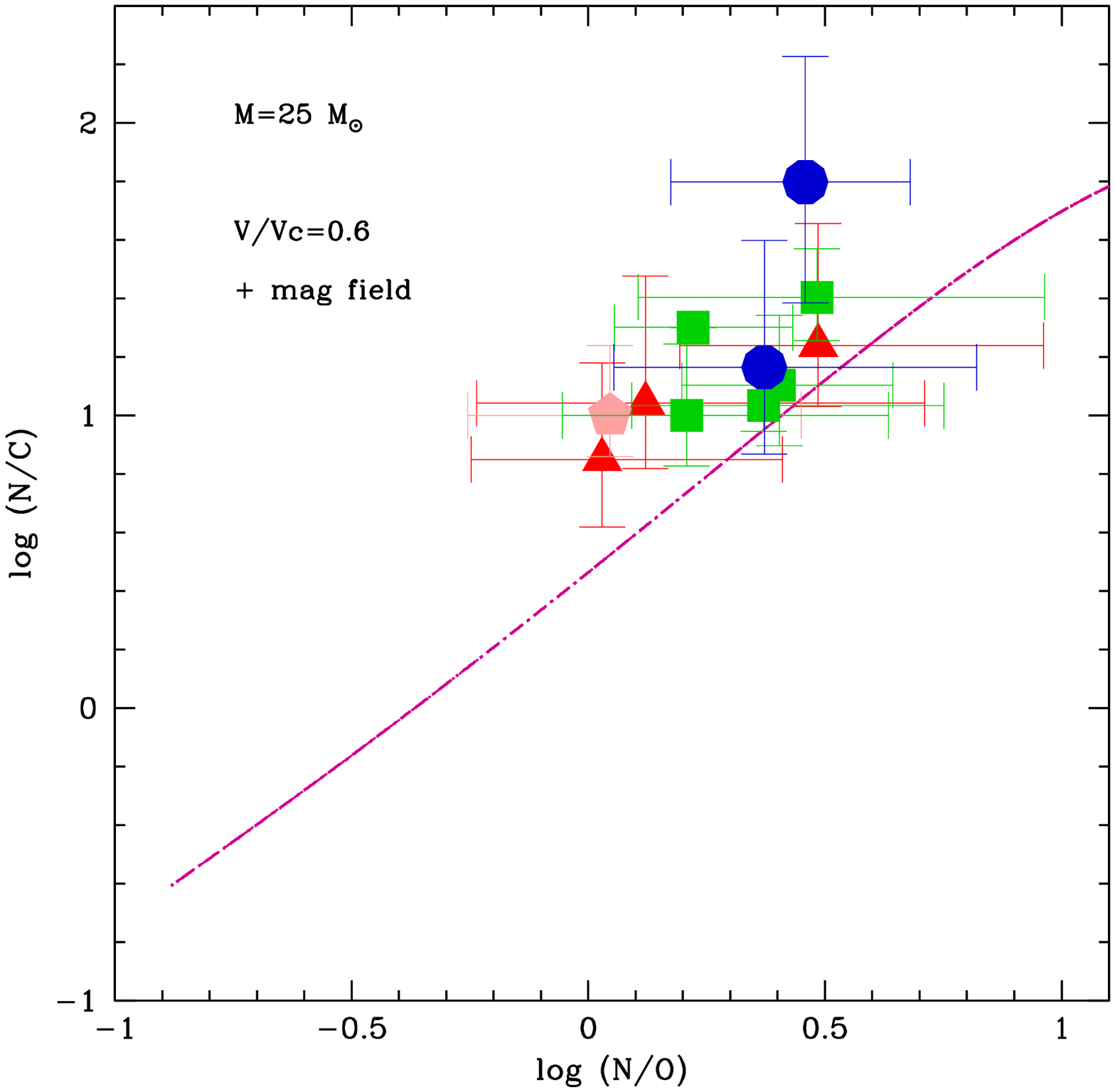}
\caption{Effect of rotation on a 25 \msun\ evolutionary model. \textit{Left:} \logg\ - log(\teff) diagram. \textit{Middle:} log(N/C) - log(N/O) diagram for models at different velocities. \textit{Right:} log(N/C) - log(N/O) diagram for a model including magnetic field and V/V$_{\rm c}$=0.6. The bold part of the tracks corresponds to the main sequence. Dots in the middle panel correspond to a surface \logg\ $= 3.5$, i.e. the minimum value for dwarfs and giants in the left-hand panel; in the right-hand panel, \logg\ is always higher than 3.5. Models for $V/V_{\rm c}=0.4$ are from \citet{ek12}. Models for higher velocities have been computed for the present study with the Geneva code until the red supergiant phase.}
\label{fig_modfast}
\end{figure*}

Scrutiny of 15-M$_\odot$ models from \citet{ge13b} suggests that
surface enrichment of CNO-cycle products can be increased
significantly (by up to a factor of 3 compared to normally rotating
stars) through adopting initial rotation rates close to the break-up
velocity.  It is therefore tempting to speculate that the nitrogen
abundances of ON stars can be attributed to enhanced efficiency of
rotational mixing in extremely fast-rotating stars.  To test this
hypothesis, new 25-\msun\ models were computed using the Geneva code,
with ratios of initial to critical equatorial velocities, V/V$_{\rm
  c}$, of 0.6 and 0.8 (see Fig.~\ref{fig_hr}).  Results are presented
in Fig.~\ref{fig_modfast}. The left-hand panel shows that increasing
the initial rotational velocity leads to a more `upward' evolution
(i.e., the effective temperature decreases less at higher
velocity). All non-supergiant ON stars are located between the
zero-age main sequence and the terminal main sequence, whatever the
initial rotation speed.  The middle panel of
Fig.~\ref{fig_modfast} shows the corresponding surface abundances; as
expected, faster rotation leads to stronger enrichment. At the end of
the main sequence, the $V/V_{\rm c} = 0.8$ model
barely reaches the region occupied by the ON stars. However, this
happens when \logg\ $< 3.5$; for the range of surface gravities of non-supergiant ON
stars (3.5--4.0) the surface enrichment is weaker. Consequently,
evolutionary models with high rotation velocities do not appear to
explain the extreme enrichment of ON stars, at least with the current formalism used to treat rotation.

The right-hand panel of Fig.~\ref{fig_modfast} shows the effect of adding magnetism to the $V/V_{\rm c} = 0.6$ model. This models includes the effects of the internal magnetic fields on the mixing of chemicals and angular momentum, according to the Tayler-Spruit dynamo theory \citep{Spruit2002,Maeder2004}. In this framework, internal magnetic fields produces a strong coupling between the core and the surface, and considerably increases the internal mixing \citep[e.g.][]{Maeder2005}. Significant enrichment, consistent with that of ON stars, can be produced while \logg\ $>$ 3.5. However, inspection of the left-hand panel of Fig.~\ref{fig_modfast} reveals that this model produces a blueward evolution and never reaches the position of the ON stars.

\begin{figure*}[]
\centering
\includegraphics[width=0.45\textwidth]{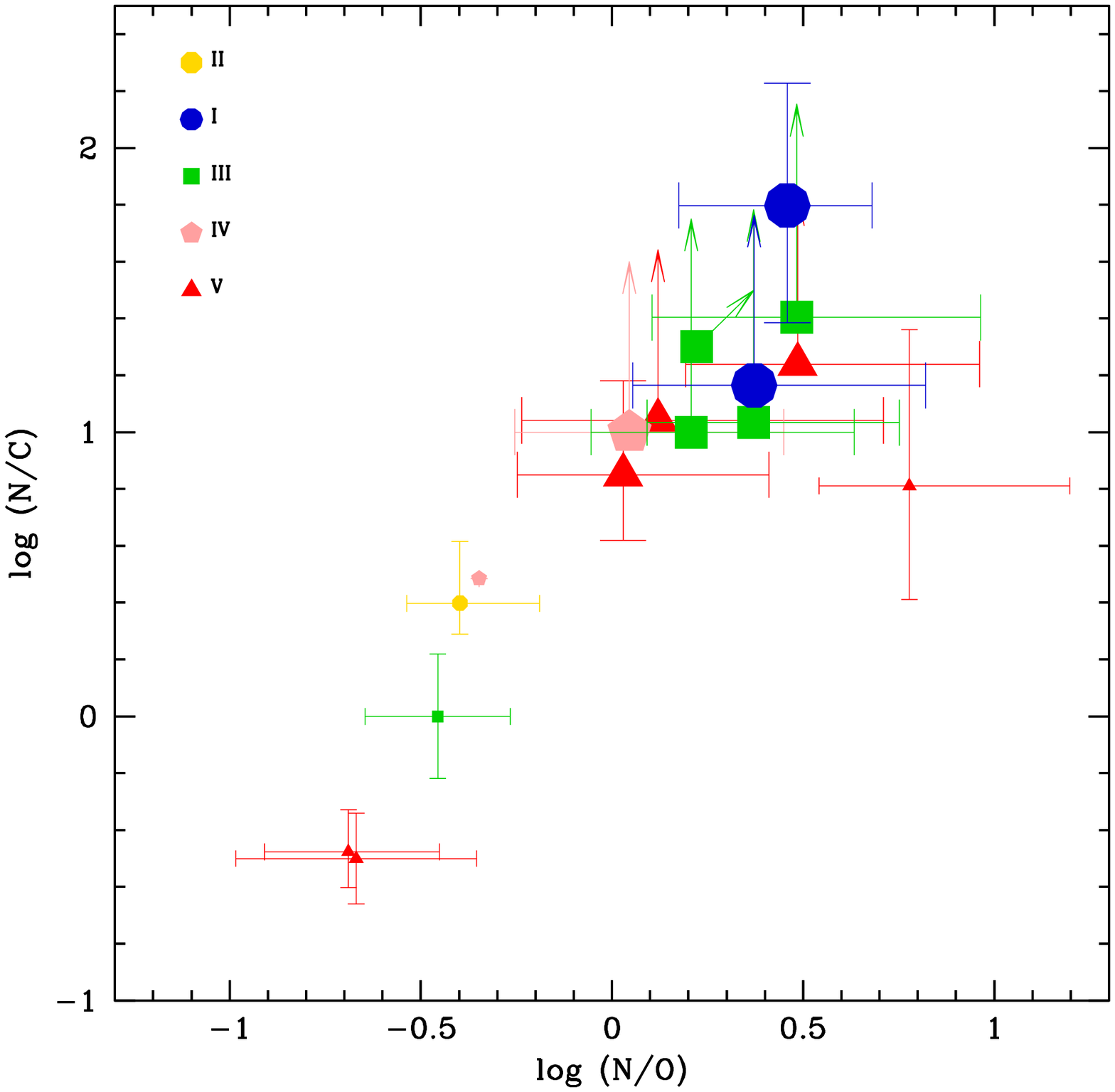}
\includegraphics[width=0.45\textwidth]{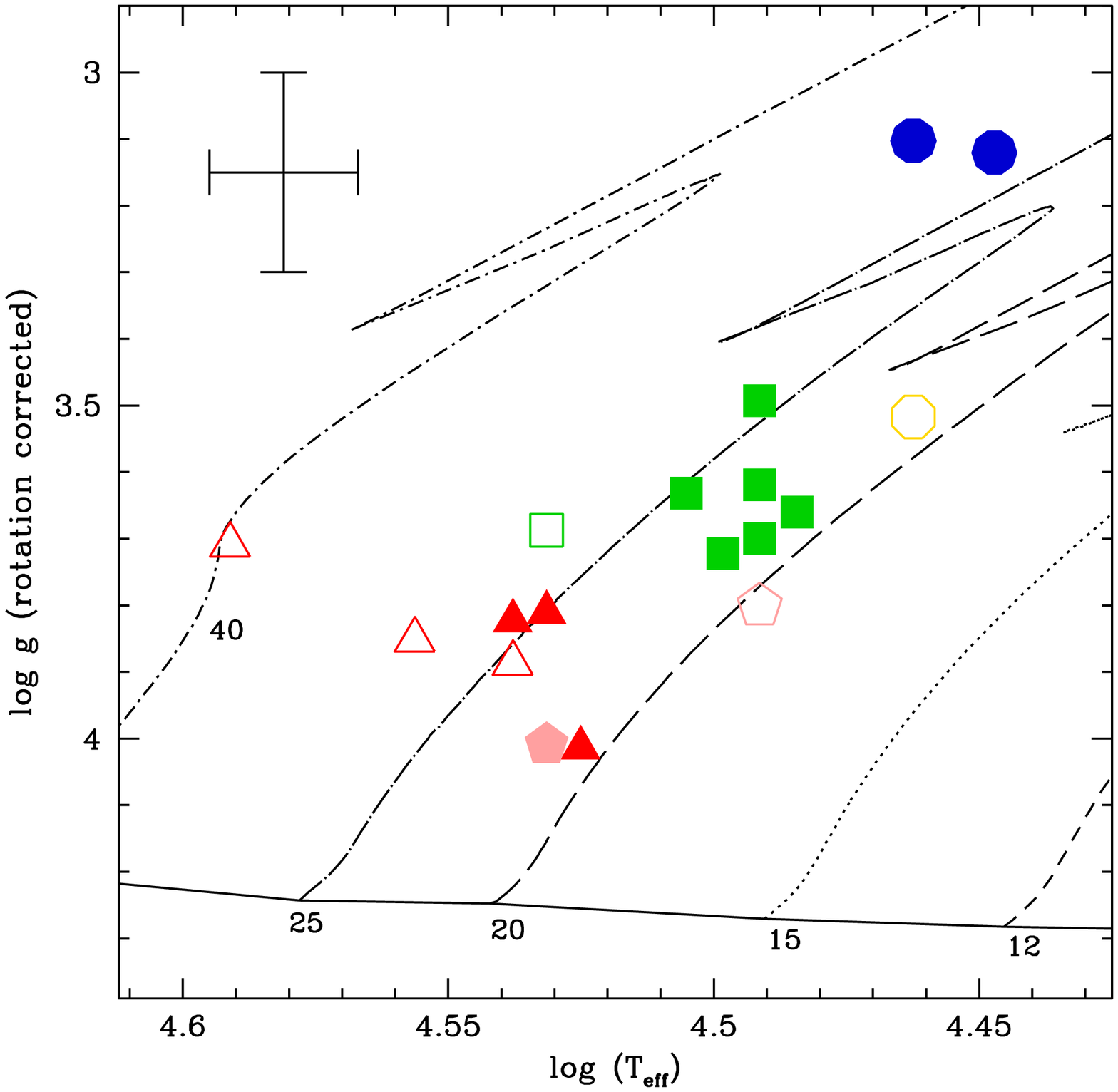}
\caption{Comparison between ON and non-ON fast rotating stars. \textit{Left}: log(N/C) vs. log(N/O) diagram. \textit{Right}: \logg--log(\teff) diagram. Large symbols are ON stars form the present study; open/small symbols are fast rotating stars (\vsini\ $>$ 250 \kms) from \citet{mimesO}. The star $\zeta$~Oph is also included: effective temperature and surface gravity are from \citet{mimesO} while surface abundances are from \citet{vil05}.}
\label{fig_fast}
\end{figure*}

Models aside, if rotational mixing \emph{is} the principal cause of
the ON phenomenon, one may wonder why not all fast-rotating O stars
are classified as ON. For example, $\zeta$~Oph has a projected
rotational velocity of about 400~\kms\ \citep[e.g.,][]{wagner09}, but
its spectral type is simply O9.2\;IVnn \citep{sota14},
where the `nn' qualifier reflects the strong rotational broadening;
that is, $\zeta$~Oph is not classified as an ON star \citep[see also][]{hs01}.
\citet{vil05} performed a quantitative analysis of its stellar parameters and surface abundances (using equivalent widths measurements and the curve-of-growth method). \citet{wagner09}, using more recent atmosphere models than Villamariz \& Herrero, refined the \teff\ and \logg\ determination. $\zeta$~Oph, together with the fast rotators of the sample of \citet{mimesO} (i.e. stars with \vsini\ $>$ 250 \kms) are shown in Fig.\ \ref{fig_fast}. 
The left-hand panel of Fig.~\ref{fig_fast} demonstrates that the ON
stars are more chemically processed than the non-ON fast rotators. The
only non-ON fast rotator with a strong enrichment is HD~192281
(O4.5\;Vn(f)), the most massive dwarf of the sample; its enrichment is
probably due to its higher mass \citep[see discussion in][]{mimesO}.

Since chemical enrichment depends not only on rotation but also on metallicity, mass and age, one may wonder if the non-ON fast rotators are less evolved or less massive than the ON stars (supposing their
metallicities to be similar since all stars are located in the Galaxy). The right-hand panel of Fig.\ \ref{fig_fast} shows that ON and non-ON stars are distributed in the same area of the \logg--\teff\ diagram. Thus, they have similar mass and age ranges. In conclusion, if ON stars appear to rotate on average faster than normal stars, not all fast rotators necessarily display the ON phenomenon. Another mechanism may be required, in addition to rotation, to produce ON stars.

\subsection{Binarity and the ON phenomenon}
\label{s_var}

\begin{figure*}[t]
\centering
\includegraphics[width=0.9\textwidth]{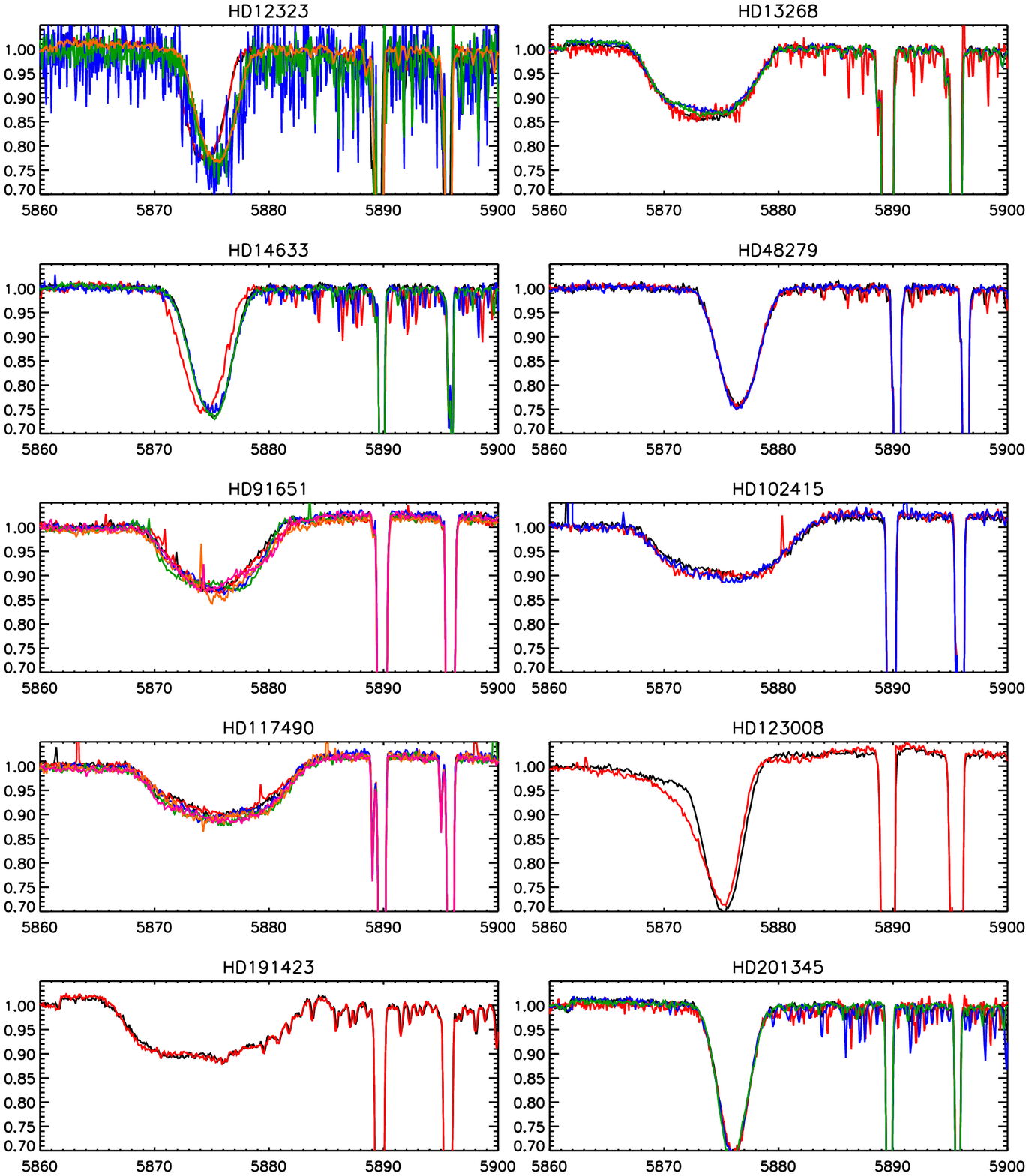}
\caption{Variability of ON stars. For each object, a selection of spectra centered on \ion{He}{i}~$\lambda$5876 is shown. The spectra are described in Table~\ref{tab_var}.}
\label{fig_var}
\end{figure*}

\citet{br78} reported that most of the ON stars known at the time of their study were radial velocity (RV) variables, but OC stars were all RV constant. \citet{walborn11} discussed radial-velocity variability and the presence of companions in the most recent ON sample. Of the two O supergiants in our study, HD~123008 is reported to be constant in radial velocity, while HD191781 may be variable. The variability status of the giants is quite uncertain, with some claims of RV variations but based on limited numbers of measurements. The nature of the tentative variability observed in the giant sample is also unclear: binarity or wind/photospheric variability? Of the five dwarfs/subgiants, HD~48279 is constant in radial velocity \citep{mahy09}, HD~12323 is a single-lined spectroscopic binary (SB1; \citealt{br78}), HD~14633 an SB1 with a low mass component and a high eccentricity \citep{boy05}, HD~102415 is reported as ``likely variable'' by \citet{walborn11} and HD~201345 has an unclear status. 

Figure~\ref{fig_var} provides more information on the line-profile and
radial-velocity variations of ten of the ON stars studied in the
present paper. The data used to build Fig.~\ref{fig_var} are
described in Table \ref{tab_var}. We chose \ion{He}{i}~$\lambda$5876
for study, as nearby telluric lines can be checked for
wavelength-calibration issues. We confirm that HD~12323 and HD~14633
are SB1, with clear RV variations. HD~48279 is very stable in our
data, nor do HD~191423 or HD~201345 show signs of variability. All
other stars examined (HD~13268, HD~91651, HD~102415, HD~117490,
HD~123008) show some degree of RV variability, the origins of which
are unclear (RV modulations, pulsations, wind variability).

The exact position in respect of binarity among ON star is thus not
completely clear. Variations in radial velocities are observed in a
large fraction of the sample, but these are unambiguously attributable
to binarity only in the few cases for which an orbital solution can be
performed (HD~12323, HD~14633) or when an SB2 spectrum is observed
(HD~89137). For the majority of the sample stars, radial
velocity-variations do not have an unambiguous origin (they may be due
to binarity or wind/photospheric variability), while a few stars'
spectra are consistent with constant radial velocities.

This raises the question of the role of binarity in the strong
chemical mixing that we have shown to be a clear-cut characteristic of
ON stars. More specifically, one may wonder if mass transfer is
responsible for this strong enrichment. In the classical mass-transfer
scenario, material processed through nucleo{\-}synthesis in the core
of a companion is dumped onto what we observe today as an ON
star. Stars with an SB1 spectrum may be consistent with this scenario;
the secondary may be a low-mass star, possibly a compact object
resulting from the evolution and supernova explosion of an initially
more massive star than the present-day primary. This initially more
massive star could have contaminated the companion that we observe
today as chemically enriched before exploding as supernova. The system
would have had to stay bound during the supernova phase to account for
the SB1 RV curve.  A mass-transfer scenario may also be applicable to
RV-constant ON stars. These stars could have been initially part of
binary systems in which mass transfer occurred before disruption in
the supernova phase.

\citet{br78} argued that no sign of current episodes of mass transfer
was observed among ON stars, especially dwarfs. Plaskett's star is a
famous binary system which is very probably in a post Roche-lobe
overflow phase \citep{linder08} and which shows strong line emission
and variability. This is not observed currently among the ON stars we
studied in the present paper, indicating that active mass transfer is
probably not important among them. HD~48279, which has very similar
parameters to the SB1 system HD~14633, does not show RV
variations. Both stars are located very close to each other in the
\logg--\teff\ diagram (Fig.\ \ref{fig_hr}), hence their surface
chemical enrichment cannot be clearly attributed to mass transfer,
given that one is in a binary system and the other apparently is not,
although we cannot exclude the possibility that HD~48279 is the result
of the merger of two stars, which could explain, at least
qualitatively, the strong mixing of CNO material.

However, mass transfer is very probably taking place in the binary
BD+36~4063 (not included in our sample). 
\citet{williams09} showed that this system contains a ON9.7Ib star
with periodic radial-velocity variations. The secondary component is
not observed, although for its estimated mass it should be seen in
high-resolution spectra. \citeauthor{williams09} argue that the ON
star is currently transferring mass to the companion, surrounding it
by a thick disk, preventing the secondary from being detected
spectroscopically. In this system, therefore, the ON star is currently
the mass donor. 

A parallel situation is encountered in the massive SB2 system LZ~Cep.
Unlike BD+36~4063, the companion in LZ~Cep is clearly detected,
excluding the presence of a thick circumstellar disk arising from
ongoing mass transfer.  \citet{lzcep} obtained multi-epoch
spectroscopy of the system and disentangled the spectra of both
components, to which they assigned spectral types of O9\;III + ON9.7\;V
(primary, secondary).  Their quantitative analysis of the individual
components' spectra showed the secondary to be significantly
nitrogen rich and carbon/oxygen poor: log(N/C) (respectively
log(N/O)) reaches 1.6 (1.3).  The secondary component therefore shows
the typical abundance patterns of ON stars, as established in the
present study. \citet{lzcep} interpreted the surface abundances of the
secondary as the result of a previous episode of mass transfer in
which the secondary was initially the most massive star.
The ON star in LZ~Cep evolved faster and transferred mass to what was
original the secondary (now seen as the primary). In this scenario,
the abundance pattern of the ON secondary star is due to the removal
of external layers during the mass-transfer phase, thereby exposing
internal layers of the star. These internal layers are more chemically
mixed and thus show CNO processed material; that is, for LZ~Cep (and
possibly BD+36~4063) the ON star would have been created because of
mass \textit{removal} and not mass \textit{accretion}.

Thus, the role of binarity (and mass transfer) in the appearance of
the ON phenomenon is not obvious. It is even less clear if we consider
once again the fast rotator $\zeta$~Oph, which is a runaway
\citep{bla61,tetz11}. Runaways may be produced by dynamical
interactions in young, dense clusters or by supernova ejection from a
binary system. Dynamical interaction in clusters usually involves at
least one binary system \citep{hb83,hoffer83,hoog01}. Consequently,
there is a high probability that $\zeta$~Oph was part of a binary
system at an earlier stage in its evolution; in particular,
\citet{hoog00} \citep[see also][]{vr96} argue that $\zeta$~Oph
resulted from a supernova kick. Whatever the evolutionary and dynamical
history of this star, it did not reach the level of enrichment seen in
ON stars. Consequently, the combination of binarity and fast rotation
(possibly acquired through tidal spin-up) does not appear to
lead inevitably to the ON phenomenon.

\section{Conclusions and final remarks}
\label{s_conc}

We have performed a spectroscopic analysis of a sample of ON stars using atmosphere models computed with the code CMFGEN. We have determined the fundamental parameters and the He, C, N, and O surface abundances, along with projected rotational velocities. The results can be summarized as follows :

\begin{itemize}

\item[$\bullet$] These late-type ON stars have initial masses in the range 20 to 25 \msun, with the two supergiants of the sample being slightly more massive.

\item[$\bullet$] The projected rotational velocities of ON stars are on average higher than those of comparison samples of normal O stars \citep{mimesO,sergio14}.

\item[$\bullet$]  ON stars are chemically enriched;  almost all
  have a helium to hydrogen number ratio larger than 0.1. All stars
  are nitrogen rich, carbon poor, and, to a lesser extent, 
  oxygen poor. The surface CNO abundances are consistent with
  nucleo{\-}synthesis predictions.

\item[$\bullet$] ON stars are more chemically mixed than morphologically normal stars of similar spectral types and luminosity classes in the sample of \citet{mimesO}.

\item[$\bullet$] Some ON stars are members of binary systems; some show radial-velocity variations of unclear origin; some are constant in radial velocity.

\item[$\bullet$] Evolutionary models including rotation are not able to reproduce the high degree of chemical mixing observed on the main sequence among ON stars.

\end{itemize}

From these results, we conclude that ON stars are O stars showing
CNO-processed material at their surface; the N/C and N/O ratios we
establish are the highest observed so far in O stars. The processed
material is the product of nucleo{\-}synthesis, but the mechanisms by
which it is brought to the surface remains unclear (rotational mixing
and binary mass transfer being the prime suspects). ON stars rotate on
average faster than normal stars, but there exist fast rotators
with similar masses and ages that are not ON stars. There are also binary and
(presumably) single stars among the ON category, as well as non-ON
fast rotators that are (probably) former members of binary systems.

\section*{Acknowledgments}

We thank John Hillier for making CMFGEN available to the community. FM and AP thank the Agence Nationale de la Recherche for financial support (grant ANR-11-JS56-0007). SS-D acknowledge funding by the Spanish Ministry of Economy and Competitiveness (MINECO) under the grants AYA2010-21697-C05-01, AYA2012-39364-C02-01, and Severo Ochoa SEV-2011-0187, and by the Canary Islands Government under grant PID2010119. C.G. acknowledges support from the European Research Council under the European Union’s Seventh Framework Programme (FP/2007-2013) / ERC Grant Agreement n. 306901. RB ackowledges support from FONDECYT Regular Project 1140076. STScI is operated by the Association of Universities for Research in Astronomy, Inc., under NASA contract NAS5-26555. We thank an anonymous referee for a timely and positive report.

\bibliographystyle{aa}
\bibliography{on}

\newpage

\begin{appendix}

\section{Additional observational information}

Table \ref{tab_var} provides information on the spectra used to investigate spectral variability and binarity in the ON stars sample. The spectra are shown in Fig.\ \ref{fig_var}. 

\begin{table*}
\begin{center}
\caption{Spectroscopic data used to study variability.} \label{tab_var}
\begin{tabular}{llll}
\hline
Star        & date of observation &   Instrument  &  Status  \\    
\hline
HD~12323    & 08 sep 2011         &   N-NOT/FIES  &  SB1 \\
            & 12 sep 2011         &   N-NOT/FIES  &  \\
            & 29 oct 2012         &   M-MERCATOR/HERMES & \\
            & 29 oct 2012         &   M-MERCATOR/HERMES & \\
            & 25 dec 2012         &   N-NOT/FIES & \\
HD~13268    & 12 jan 2011         &   N-NOT/FIES  &  variable \\
            & 26 oct 2012         &   M-MERCATOR/HERMES & \\
            & 28 jan 2013         &   N-NOT/FIES & \\
            & 29 jan 2013         &   N-NOT/FIES & \\
HD~14633    & 13 jan 2009         &   N-NOT/FIES  &  SB1 \\
            & 10 sep 2011         &   N-NOT/FIES & \\
            & 26 oct 2012         &   M-MERCATOR/HERMES & \\
            & 23 dec 2012         &   N-NOT/FIES & \\
HD~48279    & 14 jan 2011         &   N-NOT/FIES & constant \\
            & 24 dec 2012         &   N-NOT/FIES & \\
            & 25 dec 2012         &   N-NOT/FIES & \\
HD~91651    & 04 apr 2009         &   ESO2.2/FEROS & variable \\
            & 11 feb 2011         &   ESO2.2/FEROS & \\
            & 20 mar 2011         &   ESO2.2/FEROS & \\
            & 14 may 2011         &   ESO2.2/FEROS & \\
            & 15 may 2011         &   ESO2.2/FEROS & \\
            & 16 may 2011         &   ESO2.2/FEROS & \\
HD~102415   & 12 may 2008         &   ESO2.2/FEROS & variable \\
            & 16 may 2011         &   ESO2.2/FEROS & \\
            & 17 may 2011         &   ESO2.2/FEROS & \\
HD~117490   & 12 may 2008         &   ESO2.2/FEROS & variable \\
            & 13 may 2011         &   ESO2.2/FEROS & \\
            & 20 mar 2011         &   ESO2.2/FEROS & \\
            & 21 mar 2011         &   ESO2.2/FEROS & \\
            & 14 may 2011         &   ESO2.2/FEROS & \\
            & 15 may 2011         &   ESO2.2/FEROS & \\
HD~123008   & 13 may 2008         &   ESO2.2/FEROS & variable \\
            & 16 may 2011         &   ESO2.2/FEROS & \\
HD~191423   & 29 aug 2011         &   N-NOT/FIES & constant \\
            & 11 sep 2011         &   N-NOT/FIES & \\
HD~201345   & 09 sep 2010         &   N-NOT/FIES & constant \\
            & 17 jun 2011         &   M-MERCATOR/HERMES &  \\
            & 10 sep 2011         &   N-NOT/FIES &  \\
            & 25 dec 2012         &   N-NOT/FIES &  \\
\hline
\end{tabular}
\end{center}
\end{table*}

\newpage

\section{Best fits to the observed spectra}

\begin{figure}[t]
\centering
\includegraphics[width=9cm]{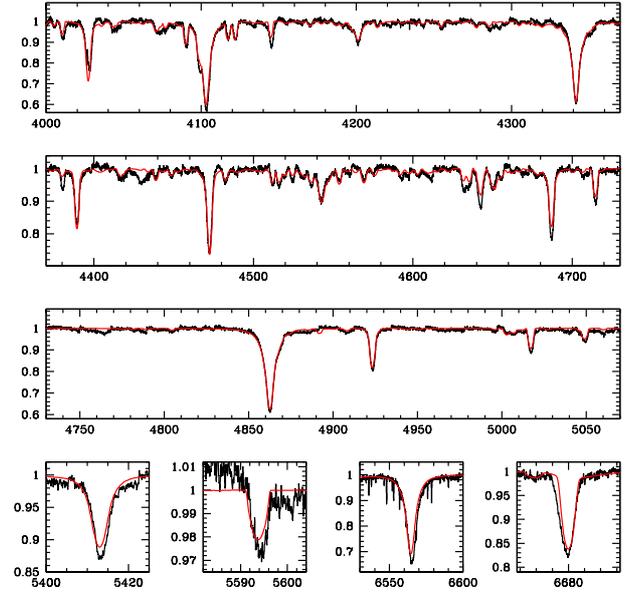}
\caption{Best fit (red) of the observed spectrum (black) of HD~12323.}
\label{fit_12323}
\end{figure}

\begin{figure}[t]
\centering
\includegraphics[width=9cm]{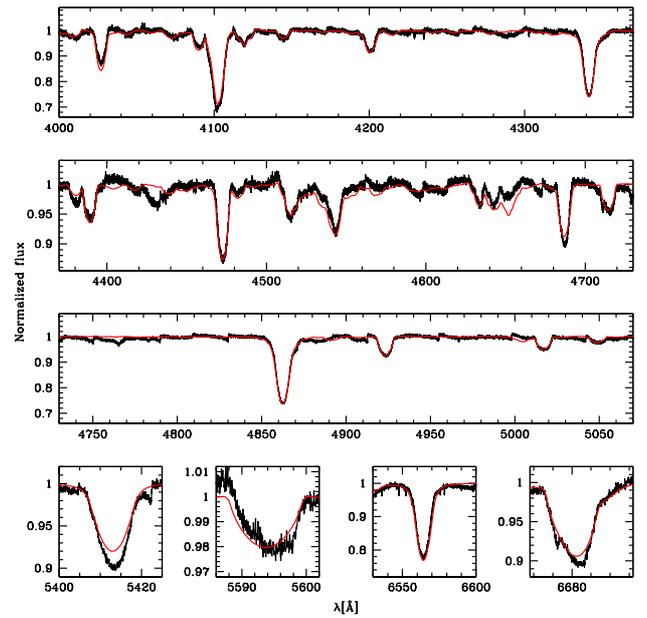}
\caption{Best fit (red) of the observed spectrum (black) of HD~13268.}
\label{fit_13268}
\end{figure}

\begin{figure}[t]
\centering
\includegraphics[width=9cm]{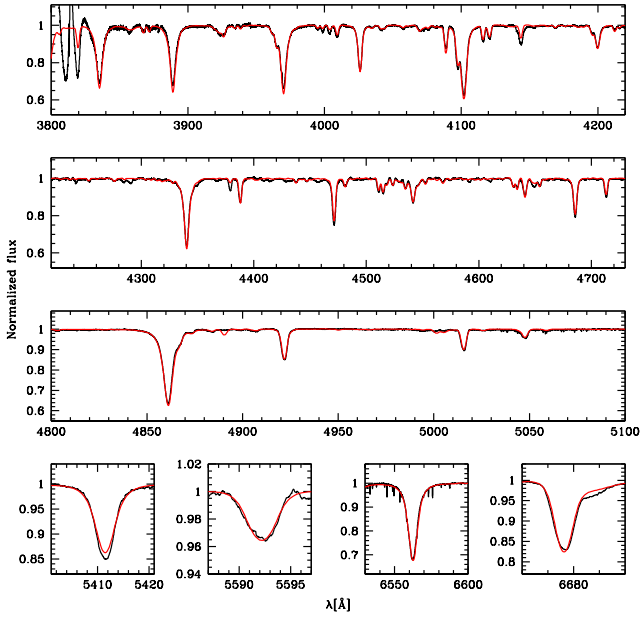}
\caption{Best fit (red) of the observed spectrum (black) of HD~14633.}
\label{fit_14633}
\end{figure}

\begin{figure}[t]
\centering
\includegraphics[width=9cm]{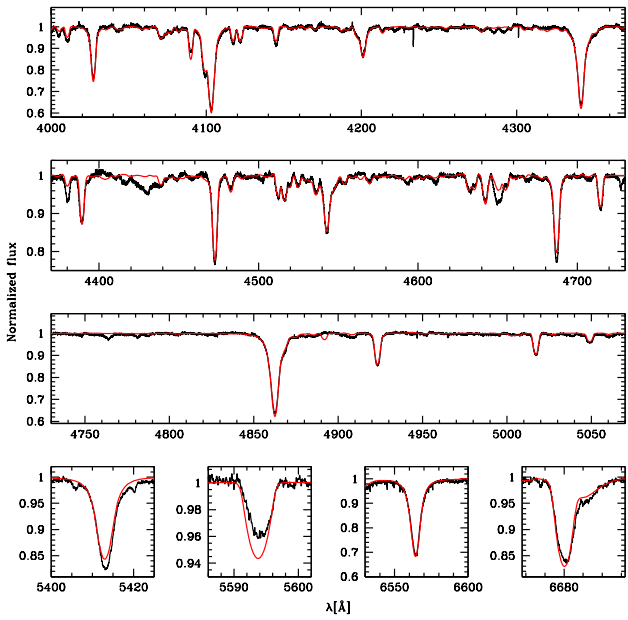}
\caption{Best fit (red) of the observed spectrum (black) of HD~48279.}
\label{fit_49279}
\end{figure}

\begin{figure}[t]
\centering
\includegraphics[width=9cm]{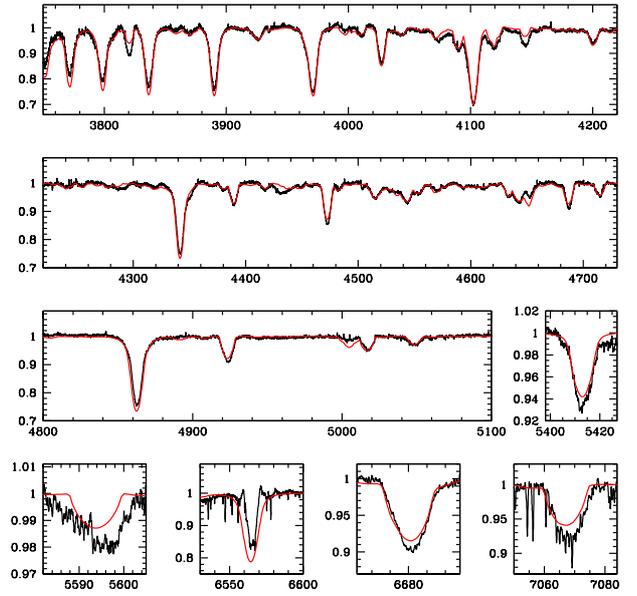}
\caption{Best fit (red) of the observed spectrum (black) of HD~91651.}
\label{fit_91651}
\end{figure}

\begin{figure}[t]
\centering
\includegraphics[width=9cm]{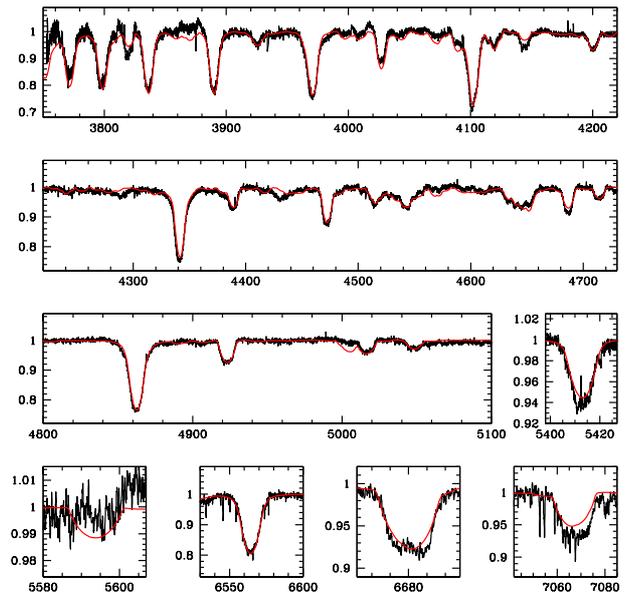}
\caption{Best fit (red) of the observed spectrum (black) of HD~102415.}
\label{fit_102415}
\end{figure}

\begin{figure}[t]
\centering
\includegraphics[width=9cm]{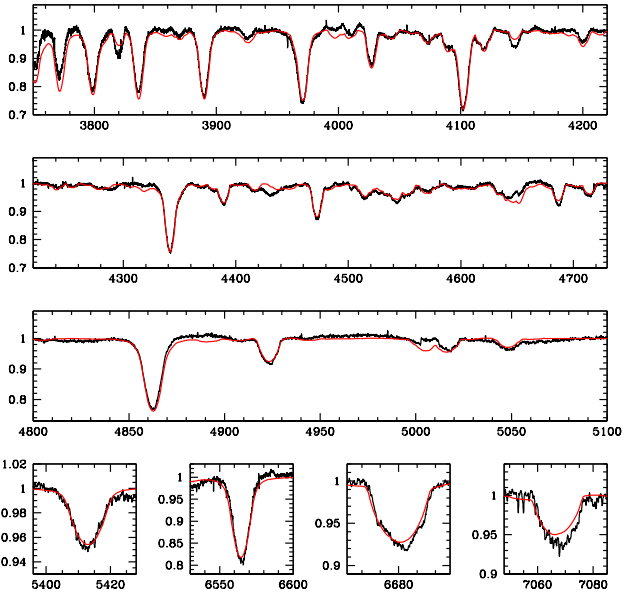}
\caption{Best fit (red) of the observed spectrum (black) of HD~117490.}
\label{fit_117490}
\end{figure}

\begin{figure}[t]
\centering
\includegraphics[width=9cm]{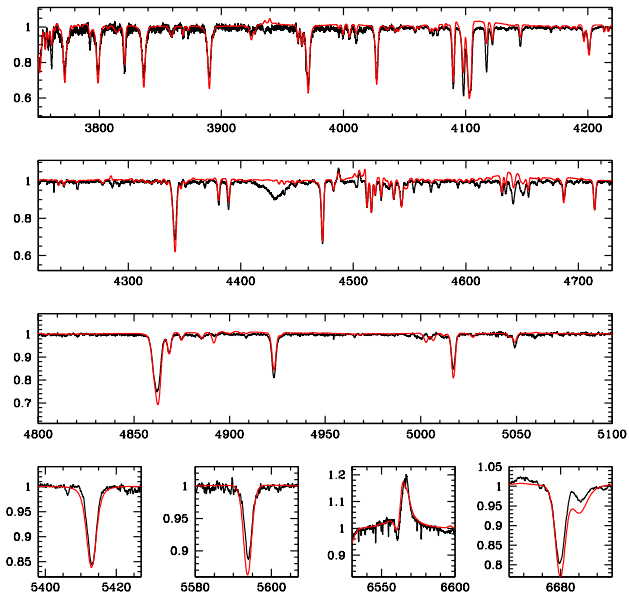}
\caption{Best fit (red) of the observed spectrum (black) of HD~123008.}
\label{fit_123008}
\end{figure}

\begin{figure}[t]
\centering
\includegraphics[width=9cm]{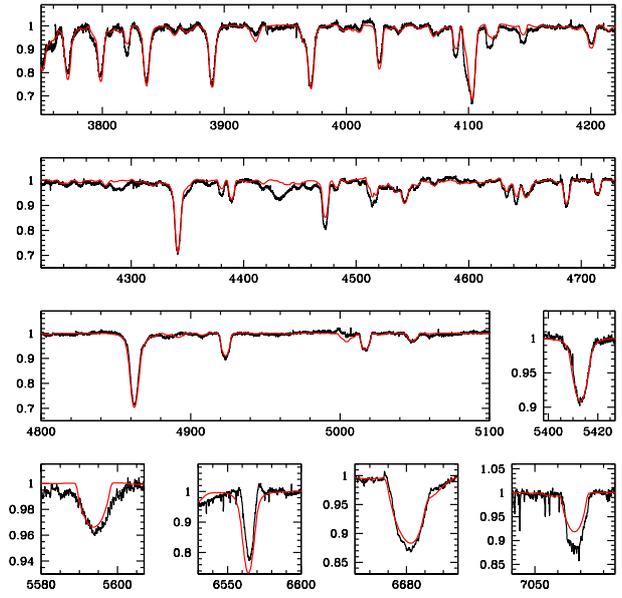}
\caption{Best fit (red) of the observed spectrum (black) of HD~150574.}
\label{fit_150574}
\end{figure}

\begin{figure}[t]
\centering
\includegraphics[width=9cm]{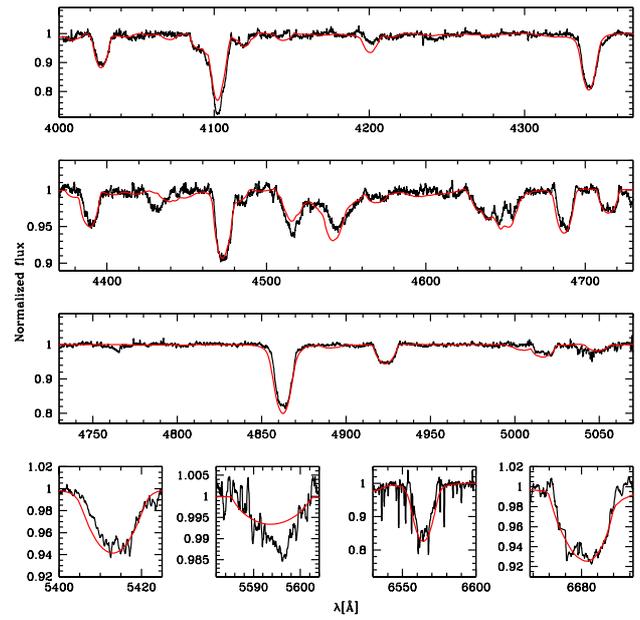}
\caption{Best fit (red) of the observed spectrum (black) of HD~191423.}
\label{fit_191423}
\end{figure}

\begin{figure}[t]
\centering
\includegraphics[width=9cm]{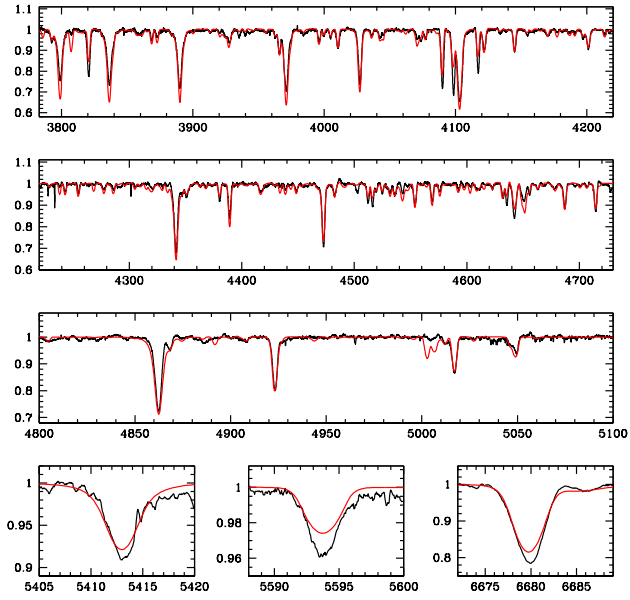}
\caption{Best fit (red) of the observed spectrum (black) of HD~191781.}
\label{fit_191781}
\end{figure}

\begin{figure}[t]
\centering
\includegraphics[width=9cm]{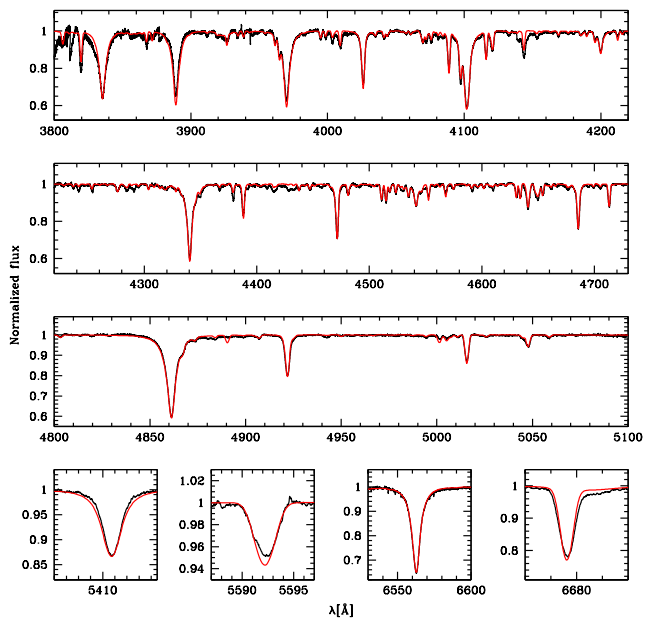}
\caption{Best fit (red) of the observed spectrum (black) of HD~201345.}
\label{fit_201345}
\end{figure}

\end{appendix}

\end{document}